\documentclass[fleqn,usenatbib]{mnras}

\usepackage[T1]{fontenc}
\usepackage{ae,aecompl}

\usepackage{ctable}

\usepackage{graphicx}	
\usepackage{amssymb}	
\usepackage{amsmath}	

\usepackage{mathptmx}
\usepackage{txfonts}

\title[Impact of GNSS on HI IM]{Potential Impact of Global Navigation Satellite Services on Total Power HI Intensity Mapping Surveys}

\author[S. E. Harper et al.]{
Stuart E. Harper,$^{1}$\thanks{E-mail: stuart.harper@manchester.ac.uk}
and Clive Dickinson
\\
$^{1}$Jodrell Bank Centre for Astrophysics, Alan Turing Building, School of Physics and Astronomy, University of Manchester, Oxford Road, Manchester M13 9PL\\
}

\date{Accepted XXX. Received YYY; in original form ZZZ}

\pubyear{2018}

\begin{document}
\label{firstpage}
\pagerange{\pageref{firstpage}--\pageref{lastpage}}
\maketitle

\begin{abstract}
Future total-power single-dish HI intensity mapping (HI IM) surveys have the potential to provide unprecedented insight into late time ($z < 1$) cosmology that are competitive with Stage IV dark energy surveys. However, redshifts between $0 < z < 0.2$ lie within the transmission bands of global navigation satellite services (GNSS), and even at higher redshifts out-of-band leakage from GNSS satellites may be problematic. We estimate the impact of GNSS satellites on future single-dish HI IM surveys using realistic estimates of both the total power and spectral structure of GNSS signals convolved with a model SKA beam. Using a model of the SKA phase one array with 200\,dishes we simulate a HI IM survey covering 30000\,sq.\,deg. of sky. We compare the integrated GNSS emission on the sky with the expected HI signal. It is found that for frequencies $> 950$\,MHz the emission from GNSS satellites will exceed the expected HI signal for all angular scales to which the SKA is sensitive when operating in single-dish mode.

\end{abstract}




\section{Introduction}

HI tomography of the Universe has the potential to provide information on the full evolution of large-scale structure (LSS) between $0 < z < 300$ \citep{Pritchard2008,Loeb2008}. The HI intensity mapping (HI IM) technique is expected to be an extremely efficient method of performing a survey of LSS and exploratory attempts are already being made with existing single-dish \citep[Parkes,][]{Pen2009,Anderson2017} \citep[GBT,][]{Chang2010,Masui2010,Masui2013,Shaw2014,Wolz2016} observatories with varying degrees of success. Attempts have also been made with interferometric observatories such as the \citet[MWA, ][]{Ewall2016} \citet[GMRT, ][]{Ghosh2011}, \citet[LOFAR,][]{patil2017} and \citet[PAPER, ][]{Ali2015}. In the near future single-dish HI IM surveys using the MeerKAT \citep{Santos2017} and ultimately the SKA-MID \citep{Santos2015} arrays as collections of single-dish telescopes are expected to provide constraints on late-time cosmology that is competitive with Stage IV dark energy surveys \citep{Bull2015}.

Unfortunately HI line emission, even when integrated over many galaxies, is comparatively faint when compared to other sources of radio emission at frequencies $< 1420$\,MHz. An obvious source of radio emission is the Galaxy, which is approximately 4$-$6 orders-of-magnitude brighter than the expected HI intensity field \citep[e.g.,][]{Alonso2015}. As well as astrophysical foregrounds other contaminants to the data can originate from the systematics within the receiver system such as correlated $1/f$ noise \citep{BigotSazy2015,Harper2017}, but also from the man-made radio environment surrounding the observatory. One extremely problematic form of man-made radio frequency interference (RFI) for radio astronomy comes from satellites within the global navigation satellite service (GNSS). 

The GNSS is composed of several constellations such as the United States funded Global Positioning System (GPS), the Russian Global Navigation Satellite System (GLONASS), the European Galileo constellation, the Chinese BeiDou system, and several regional or support constellations run by India, Japan and Europe \citep{hofmann2007}. In total there are approximately 60 GNSS satellites in orbit at present, which is set to increase to 120 by 2030 \citep{gao2012}, and each one has a flux density when observed in-band comparable to the quiet Sun ($\ge 10^{6}$\,Jy).

The prevalence and brightness of transmissions from GNSS satellites will have a significant impact on any future \textit{total power} single-dish HI IM survey such as those planned with the SKA \citep{Santos2015}, MeerKAT \citep{Santos2017}, BINGO \citep{Battye2013} or FAST \citep{BigotSazy2016}. As single-dish observatories have no method of discriminating between emission picked up within the main beam or the sidelobes of the telescope, extremely bright sources of emission such as GNSS satellites will be imprinted into the data. The impact of GNSS satellites on future single-dish HI IM surveys will depend on the magnitude of the GNSS fluctuations, picked up within sidelobes, compare with the expected fluctuation scale of the HI intensity signal. However, for \textit{interferometric} surveys the impact of GNSS emission should be significantly reduced because the satellite signals will not be perfectly correlated between dishes and GNSS emission is expected to be more problematic on large-scales to which interferometers are not sensitive.

There are a couple of obvious solutions to reducing the impact of the GNSS transmissions. One is to design a bespoke telescope for HI IM that has very low sidelobes. The other mitigation method is to simply not observe \textit{close} to the GNSS transmission bands. However, it should be stressed that GNSS transmissions are not necessarily constrained to the frequency bands allocated to each constellation. GNSS frequency allocations are set by the International Telecommunications Union (ITU) \citep{ITU2004}, who also provide guidelines for the maximum \textit{out-of-band} power that is allowed to leak out of these allocations \citep{ITU2015}. However, the out-of-band GNSS transmissions can still be extremely bright relative to astrophysical and cosmological sources of emission, even when only being measured in the sidelobes of a telescope. Determining exactly how bright these out-of-band transmissions are within the sidelobes of a model SKA-MID telescope when compared with the expected HI intensity field is the goal of this work.

The rest of the paper is set out as follows: In Sec.~\ref{sec:simulations} a description of the methods used to simulate the telescope beam model, observing strategy and satellite transmissions is described. Sec.~\ref{sec:Results} then takes these simulations and determines the frequency range and spatial scales that are expected to be highly contaminated by GNSS satellites. Finally Sec.~\ref{sec:conclusion} will bring together some thoughts and the main conclusions of this work.

\section{Simulations}\label{sec:simulations}

In this section the pipeline used to perform the end-to-end simulations of the expected RFI from GNSS transmissions is described. The code used to perform these simulations is the same as that described in \citet{Harper2017} and the reader is directed there for further details about the pipeline. This section will describe only the relevant parts of the pipeline for this work, and the expansions required to included GNSS transmissions.

The basic receiver model that is used to describe the time-ordered-data (TOD) outputs from the simulation pipeline is
\begin{equation}\label{eqn:rec1}
	d(t, \nu) = G_r(\nu) B_r(\Omega-\Omega_0, \nu) \ast T_\mathrm{sats}(\Omega, \nu),
\end{equation}
where the TOD vector ($d$) is a function of both time and frequency. The term $T_\mathrm{sats}(\Omega, \nu)$ is a set of delta functions describing the antenna temperature contributions of each satellite at position $\Omega$ in the sky at a given time and frequency $\nu$. The calculation of $T_\mathrm{sats}$ is described in detail in Sec.~\ref{sec:gnss}. The terms $G_r(\nu) B_r(\Omega-\Omega_0, \nu)$ describe the gain of the beam in any given direction $\Omega$ at a given time relative to the main beam axis $\Omega_0$. The beam model is described in detail in Sec.~\ref{sec:beam}. The final TOD $d$ is generated by convolving the beam model with every satellite at each time interval. There is no noise contribution included in the end-to-end simulations, therefore no additional system temperature contribution is included in Eqn.~\ref{eqn:rec1}. However, later in Sec.~\ref{sec:Results} the simulated GNSS transmissions are compared to an analytical model of the noise that assumes a model SKA phase one array with 200\,dishes.

\subsection{Survey Design}\label{sec:surveydesign}

\ctable[
  caption = Input parameters describing the simulated telescope and survey designs.,
  label = table:survey,
  width = 0.45\textwidth
]{l|c|c}{
	}{\FL Description & Parameter & Value \ML
  Dish Diameter & $D_{\mathrm{dish}}$ &  13.5\,m \NN
 Receiver Temperature & $T_\mathrm{rx}$ & 20\,K \NN
 No. Channels & $N_{\nu}$ & 130 \NN 
 Bandwidth & $\Delta \nu$ & $750 < \nu < 1400$\,MHz  \NN
 Channel width & $\delta \nu$ & 5\,MHz \NN
 Latitude & $\theta$ & $-$30.71$^{\circ}$ \NN
 Longitude & $\phi$ & 21.45$^{\circ}$ \NN
 Sample Rate & $f_\mathrm{sr}$ & 4\,Hz \NN \hline
 \textbf{Drift Scan Mode} & & \LL
 Declinations & $\delta$ & $-$10, $-$40, $-$70$^\circ$ \NN
 Integration Time & $T_\mathrm{obs}$ & 1\,day \NN \hline
 \textbf{Survey Scan Mode} & & \LL
 Elevation & E & 30$^{\circ}$ \NN
 Slew Speed & $v_{\mathrm{t}}$ & $1^{\circ}$\,s$^{-1}$ \NN
 Integration Time & $T_\mathrm{obs}$ & up to 90\,days \NN
 }
 
The goal of these simulations is to determine the impact of GNSS satellites on the SKA HI IM survey described in \citet{Harper2017}, which models 200\,dishes, observing of 30000\,sq.\,deg. for a total observing time of 30\,days. Similar SKA HI IM survey designs are described in \citet{Santos2015} and \citet{Bull2015}. However, since the GNSS signal in these simulations will repeat on 12\,hour time scales (in reality perturbations in satellites orbits will induce long timescale drifts) and all the SKA dishes are located within a small angular separation on the Earth, all 200 SKA dishes will all see the same integrated GNSS sky signal. Therefore, these simulations only simulate a single SKA dish located at the centre of the SKA array.

The observing strategy assumed for an SKA HI IM survey will be to slew all 200\,dishes continuously in azimuth at a single fixed elevation of 30$^{\circ}$. This will be referred to as the survey scan mode. This observing strategy relies on the sidereal motion of the sky to sweep out the full sky coverage of 30000\,sq.\,deg. covering all declinations $< 10^{\circ}$. This observing strategy will allow us to map the temporal fluctuations of the GNSS signals onto spatial modes and evaluate how the GNSS sky signal integrates with time and is spatially distributed. Observing times between 3 and 90\,days are used to evaluate how the power spectrum of the GNSS fluctuations change with time.

To investigate the properties of the GNSS transmissions at the TOD level a separate drift scan mode observing strategy is used instead of the survey scan mode described above. A drift scan simply involves pointing the telescope at a fixed azimuth and elevation, which will be associated with some declination on the celestial sphere, and using the natural rotation of the Earth to scan the sky at the sidereal rate. Fixed drift scans at three declinations were chosen: $-10^{\circ}$, $-40^{\circ}$ and $-70^{\circ}$. Since the density of GNSS satellites decreases away from the celestial equator, observing several declinations allows for the power spectra of the GNSS signals to be investigated without any mixing of spatial scales, and determine the fraction of time satellites will spend within any given angular separation from the main beam axis. 
 
The simulated receiver is assumed to have 5\,MHz frequency resolution over the frequency range of $750 < \nu < 1400$\,MHz. This frequency range extends to slightly lower frequencies than the lowest SKA Band 2 frequency ($\nu \approx 950$\,MHz) in order to explore the impact of GNSS transmissions across the full range of frequencies considered by single-dish HI IM. The choice of channel width in these simulations will have little impact, a wider channel width will act to smooth the GNSS frequency spectrum, but should have a minimal overall impact on the temporal or spatial fluctuations of the satellites relative to the HI intensity signal. A sample rate of 4\,Hz is adopted, which achieves a nyquist sampling of the main beam when observing at elevation 30$^\circ$ at a rate of 1$^\circ$\,s$^{-1}$. There is a flat receiver temperature of $T_\mathrm{rx} = 20$\,K, which is only used when estimating the noise power spectra discussed in Sec.~\ref{sec:Results} and not used in the GNSS simulations. A summary of the two observing strategies and the properties of the simulated receiver is given in Table~\ref{table:survey}.
 
\subsection{Telescope Beam Model}\label{sec:beam}

We are interested in the magnitude of the fluctuations in the receiver system temperature due to the GNSS satellite transmissions being picked up in the far sidelobes of the MeerKAT or SKA beam patterns. Estimating the magnitude of the fluctuations requires a reasonable beam model of the far sidelobe structure. An estimated beam model can be generated through the transformation of a gaussian tapered illumination function over a circular aperture. The tapered illumination function is defined as \citep{Wilson2009}
\begin{equation}
	E(\rho) = e^{-0.5 \left(\frac{\rho}{\sigma} \right)^2}
\end{equation}
where $\sigma$ is the width of the gaussian taper and $\rho$ is the number of wavelengths across the aperture
\begin{equation}
	\rho = \frac{D}{2 \lambda},
\end{equation}
with $D$ being the diameter of the dish and $\lambda$ being the wavelength.

Using the simplification of azimuthal beam symmetry the normalised beam pattern can be computed using the following integrals
\begin{equation}\label{eqn:BeamPattern}
	B_r(\theta) = \left| \frac{ \int E(\rho) j_0(\sin(\theta) \rho) \rho \mathrm{d}\rho }{\int E(\rho)\rho \mathrm{d}\rho }  \right|^2
\end{equation}
where $j_0$ is a zeroth order bessel function. The forward gain of the beam is calculated by taking the ratio of the area of a sphere over the integral of Eqn.~\ref{eqn:BeamPattern} for all co-latitudes, which results in
\begin{equation}\label{eqn:gain}
	G_r = \frac{4 \pi}{2 \pi \int B_r(\theta)\sin(\theta)\mathrm{d}\theta},
\end{equation}
the gain of the beam models derived between 750 and 1400\,MHz using this method is approximately $40 < G < 46$\,dBi.

For many results presented in Sec.~\ref{sec:Results}, Eqn.~\ref{eqn:BeamPattern} is modified to remove the main beam response out to a given radius. This is done to excise satellites that approach within a certain radius of the main beam axis without having to discard the emission of other satellites during this interval. A sine filter is applied in co-latitude, defined as
\begin{equation}\label{eqn:beamfilter}
	W(\theta) =
		\begin{cases}
			0, & \text{if}\ \theta-\theta_c < -\frac{\delta\theta}{4} \\
			1, & \text{if}\ \theta+\theta_c >  \frac{\delta\theta}{4} \\
			\frac{1}{2}\sin(2 \pi \left[\theta-\theta_c  \right]/\delta\theta ) + \frac{1}{2}, & \text{otherwise} ,
		\end{cases}
\end{equation}
where $\delta \theta$ is an arbitrary width for the filter, given here the value of $\delta\theta = 1^\circ$, and $\theta_c$ is the angular scale at which half the beam power is filtered.

\begin{figure}
\centering 
\includegraphics[width=0.5\textwidth]{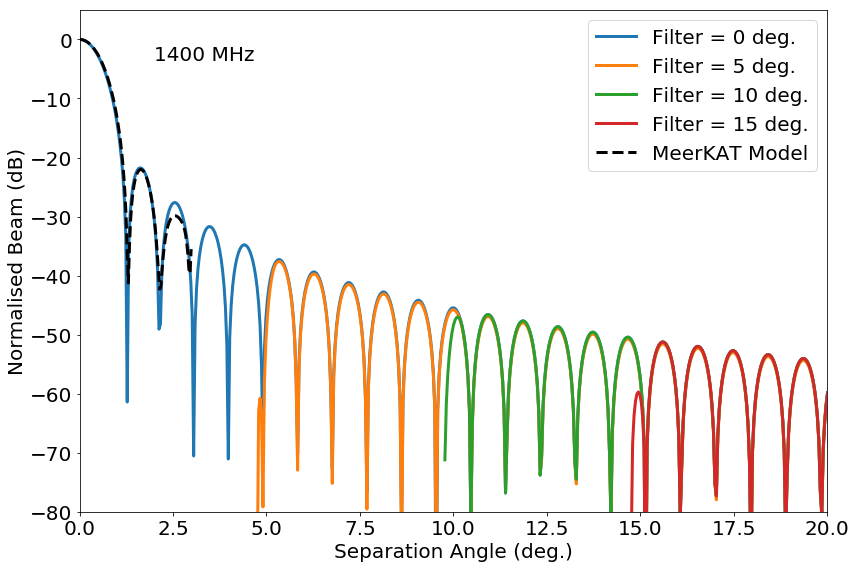}
\caption{Beam models generated assuming the transformation of the aperture illumination function as described by Eqn.~\ref{eqn:BeamPattern} at 1400\,MHz. The \textit{black dashed} line shows the predicted MeerKAT model derived using the \texttt{EIDOS}\protect\footnotemark package provided by the MeerKAT team (\textit{priv. comm.}). The different colours represent the main beam filtered models for 5$^\circ$, 10$^\circ$ and 15$^\circ$ separations.}\label{fig:Tele1}
\end{figure}
\footnotetext{\url{https://github.com/kmbasad/eidos}}

Fig.~\ref{fig:Tele1} shows the normalised beam models generated using Eqn.~\ref{eqn:BeamPattern} at 1400\,MHz. The models generated using the aperture illumination function recreate the main lobe and first sidelobe reasonably well at 1400\,MHz when compared to a measured beam model provided by the MeerKAT collaboration. The model used in this work appears to over estimate the second sidelobe. However, we are largely concerned with giving a rough guide to the expected fluctuation level. The real beam is also expected to have significantly more complex far sidelobe structure that may increase the real far sidelobe fluctuations \citep{davidson2013} that this simple beam model does not capture. The main lobe filtered beam models apply Eqn.~\ref{eqn:beamfilter} at $\theta_c=5^\circ$, 10$^\circ$ and 15$^\circ$ angular separations.

\subsection{GNSS Satellites}\label{sec:gnss}

These simulations only consider the GPS, Galileo and GLONASS constellations, other constellations such as BeiDou or regional GNSS constellations are neglected, as these are expected to have a minimal contribution at present. Similarly, GNSS support services such as the European Geostationary Navigation Overlay Service (EGNOS), or other geostationary GNSS satellites are neglected in this analysis. Geostationary satellites can contribute a significant RFI component, but are more easily avoided by simply not observing at low absolute declinations. Sec.~\ref{sec:orbits} will describe the model used to translate satellite two-line element vectors into the instantaneous telescope frame, and Sec.~\ref{sec:bright} will describe the model of the received power of the satellites, and the frequency structure of each GNSS satellite.

\subsubsection{Orbits}\label{sec:orbits}

GNSS satellite constellations are constrained to fixed orbital planes. The modern GPS network is constrained to 6-orbital planes, with four satellites per plane and 24 satellites in total. The mean altitude of the satellites is 20200\,km, giving each satellite an orbital period of approximately 12\,sidereal hours. The Russian GLONASS constellation has a slightly lower orbital altitude of 19100\,km, with 24 satellites distributed over just 3-orbital planes. Finally, the European Galileo constellation has an orbital altitude of 29600\,km, with currently 20 of the planned 30 satellites in orbit distributed over 3-orbital planes.

The orbits of each satellite were calculated using publicly available two-line element (TLE) vectors from \url{www.celestrak.com}. As we are interested in the movement of GNSS satellites within the far sidelobes of the observing telescope, a method for transforming the TLEs into the local horizon frame of the observing telescope is required. To perform the transformations we used the methods described in \citet{green1985}. Appendix~\ref{sec:app1} describes these transformations in detail.

\begin{figure}
\centering 
\includegraphics[width=0.5\textwidth]{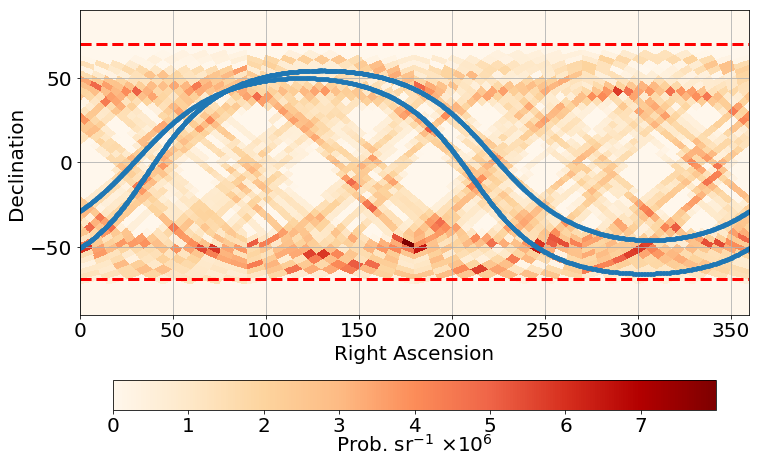}
\includegraphics[width=0.5\textwidth]{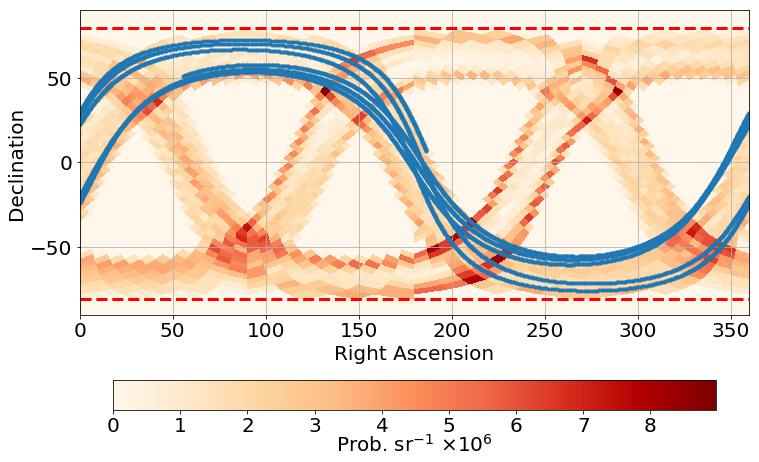}
\includegraphics[width=0.5\textwidth]{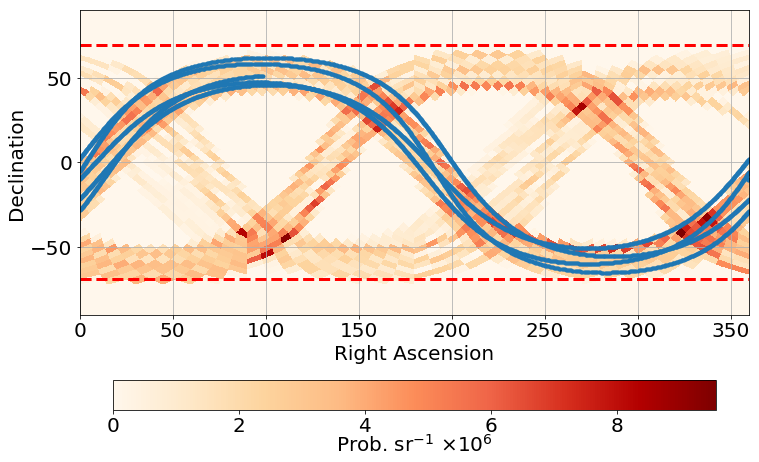}
\caption{Probability distributions of satellites across the sky per steradian for an observer at the SKA observatory. The \textit{top} plot shows the distribution of the GPS constellation, the \textit{middle} shows the GLONASS constellation, and the \textit{bottom} shows the Galileo constellation. The blue lines represents the tracks across the celestial sky of a single satellite within each constellation over a 90\,day period. The \textit{red dashed} lines show the maximum and minimum declinations of the satellites.   }\label{fig:Orbits1}
\end{figure}

Fig.~\ref{fig:Orbits1} shows the 2D probability distributions of the GPS, GLONASS and Galileo constellations across the sky. As the satellites are constrained to non-rotating orbital planes, the satellites occupy approximately fixed strips across the celestial sky. The satellites in the GPS constellations have an orbital period of 12 sidereal hours, which means an observer will see the satellite pass over twice each day. However, since the sky will rotate 12 hours between transits, each GPS satellite traces out two tracks on the celestial sphere as shown in the figure. GLONASS and Galileo have slightly shorter and longer orbital periods than GPS respectively, and therefore produce far more complex transits across the celestial sky as shown in the bottom two plots of Fig.~\ref{fig:Orbits1}. However, satellites in either constellation are still constrained to fixed strips on the celestial sky. 

From Fig.~\ref{fig:Orbits1} it can be seen that there are some regions of the sky that are more optimal to observe than others. The most clear regions of the sky are the North and South celestial poles, which never have a satellite approach closer than 25$^{\circ}$. There are then six large clean regions that are not coincident with the orbital planes of the GLONASS or Galileo constellations. However, the large number of GPS orbital planes means that there are few regions of the celestial sphere that are not, at some point, coincident with a GPS satellite.

In reality GNSS orbits are not as stable as they are modelled here, and satellites must periodically make small orbital corrections to account for being acted upon by many gravitational tidal forces. For this reason, the paths of the satellites across the sky will be more complex, especially the GPS satellite orbits. However, the satellites will still be constrained to the strips shown in Fig.~\ref{fig:Orbits1}.

\subsubsection{Satellite Brightness}\label{sec:bright}

We are interested in this paper in demonstrating the impact of the integrated GNSS satellite signal when measured within the sidelobes of the telescope and when observing at frequencies outside of the allocated GNSS bands. Sec.~\ref{sec:beam} and Sec.~\ref{sec:orbits} are two principle tools required for calculating the fluctuations of the GNSS transmissions within the sidelobes, the final tool required is a model of the frequency spectrum of GNSS signals so that the brightness of the satellites can be determined at any given frequency.

The basic GNSS transmission is a carrier frequency that is encoded using a phase modulated signal. In the most simple case, the phase will switch between $\pm \pi$, which encodes a square wave into the carrier. This form of modulation is commonly referred to as binary phase-shift keying (BPSK), and results in a power spectral density (PSD) distribution described by a simple sinc$^2$ function as \citep{hofmann2007}
\begin{equation}\label{eqn:bright1}
	S(\nu, m)_\mathrm{BPSK} = \frac{\mathrm{sinc}^2(\pi \left[\nu - \nu_c \right]/ m \nu_0)}{ m \nu_0 } ,
\end{equation}
where $\nu_c$ is the carrier frequency, and $m \nu_0$ is the frequency of the encoded rectangular phase-shift signal, referred to as the chip rate. Several GNSS services use more complex multiplexing modulation methods, which result in more complex PSD distribution models. These more complex PSDs are described in Appendix~\ref{sec:app2}.

GNSS transmissions from different constellations have different frequency allocations. For example, the GPS allocations are called L1, L2 and L5 in order of decreasing frequency. The GLONASS allocations are G1, G2 and G3, and the Galileo allocations are E1, E5 and E6. Usually two or three services are provided within each allocation per constellation that are referred to as precision (P), coarse-acquisition (C/A) and military (M) codes. Further, GPS and GLONASS satellites are separated into \textit{blocks}. Each block may only transmit a subset of services available within a given constellation and will even have differences in the power output of any given service. Table~\ref{table:power} provides a summary of satellite service allocations used in these simulations that includes frequency coverage, typical power outputs and expected PSD responses.

To calculate the received power at the Earth of GNSS transmissions requires estimates of the output power and antenna gains of each satellite per service. Power measurements are taken from observations provided in \citet{Steigenberger2017}\footnote{\citet{Steigenberger2017} provides a table of measured service powers for satellites in the GPS constellation at \url{acc.igs.org/orbits/thrust-power.txt}.}. The gain of the satellite antennas are assumed to be constant, with the GPS gain set at 13.5\,dBi taken from the Lockheed-Martin design specifications\footnote{\url{https://www.lockheedmartin.com/us/products/gps/gps-publications.html}}, 15\,dBi for the Galileo dishes based on measurements in \citet{Montesano2007}, and since no publicly available GLONASS beam model is available we adopt a nominal value of 13.5\,dBi as in \citet{Steigenberger2017}. A summary of gain antennas per satellite are provided in Table~\ref{table:power}. 

\begin{figure*}
\centering 
\includegraphics[width=0.95\textwidth]{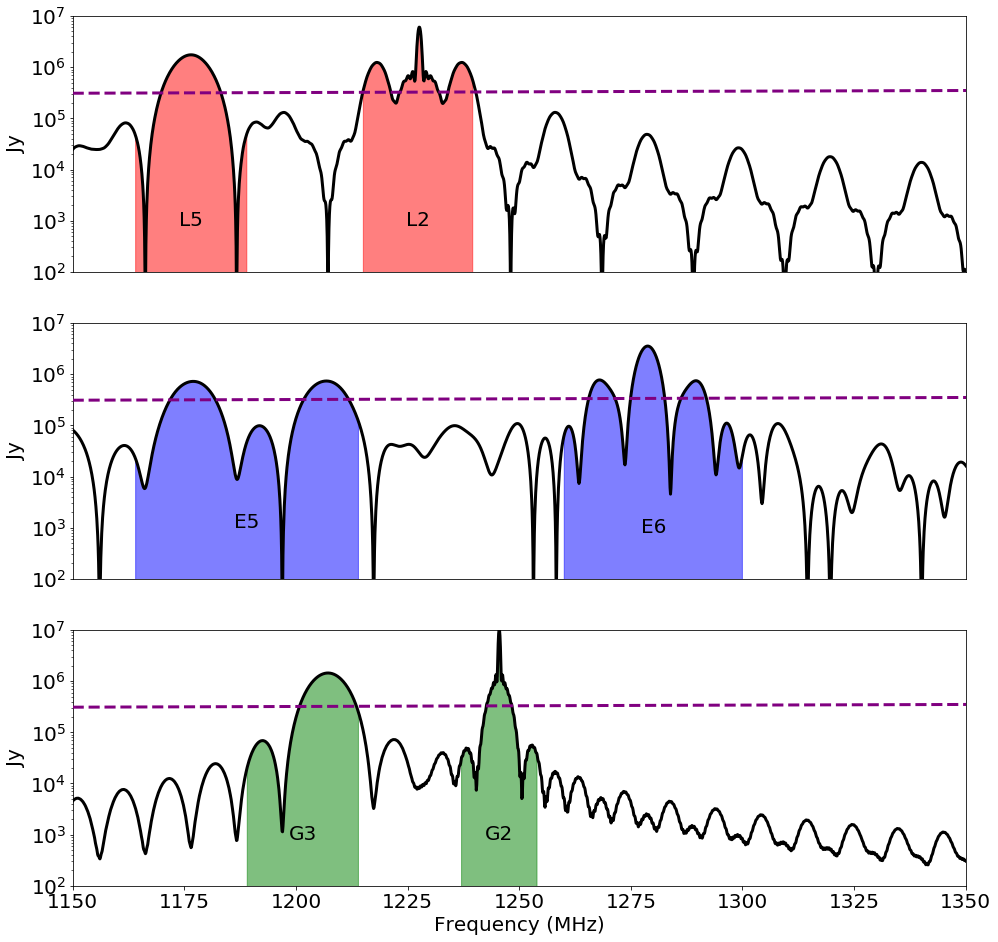}
\caption{Typical spectral energy distribution as measured from the Earth of GNSS transmissions at frequencies less than 1410\,MHz. The \textit{top} plot shows the SED for GPS, the \textit{middle} plot shows Galileo, and the \textit{bottom} shows GLONASS. Highlighted regions in the SEDs represent the nominal frequency allocations for each service and service designation. GPS services are highlighted in \textit{red}, Galileo in \textit{blue} and GLONASS in \textit{green}. Unhighlighted regions in the SED are the predicted out-of-band transmissions. The \textit{dashed purple} line shows the expected integrated flux density of the quiet Sun for reference.}\label{fig:Bright1}
\end{figure*}

The received power from each satellite is estimated using
\begin{equation}\label{eqn:satpower}
	P_\mathrm{sat} = \frac{G_t P_t}{4\pi r^2} ,
\end{equation}
where $G_t$ is the transmitter antenna gain, $P_t$ is the transmitted power and $r$ is the line-of-sight distance between the observer and satellite. Fig.~\ref{fig:Bright1} shows the integrated flux density ($S_\mathrm{sat} \approx P_\mathrm{sat}/\delta \nu$) of the services provided by each satellite constellation assuming a channel width of $\delta \nu = 1$\,MHz. For comparison, the figure shows how the expected flux density measured from the quiet Sun \citep{Hafez2014} is comparable to the expected flux density of the satellites when measured in band. However, unlike the Sun there are always at least 30 GNSS satellites above the horizon at any one time. The highlighted regions of Fig.~\ref{fig:Bright1} represent the nominal frequency allocations of each GNSS service. The flux density measured from GNSS satellites in the unhighlighted regions are still far brighter than any typical astrophysical source that might be observed at similar frequencies.

To calculate the receiver system temperature contribution of all the GNSS satellite services combined requires including the beam model of Sec.~\ref{sec:beam} and the satellite orbit model described in Sec.~\ref{sec:orbits}. The brightness temperature of one satellite at a given time and frequency is calculated using
\begin{equation}\label{eqn:oneSat} 
  T'_\mathrm{sats}(\Omega, \nu) = G_r(\nu) B_r(\nu, \Omega_0-\Omega)  \frac{P_\mathrm{sat}}{k_b \delta \nu} \frac{c^2}{4\pi \nu^2} ,
\end{equation}
where $P_\mathrm{sat}$ is the received satellite power from Eqn.~\ref{eqn:satpower}, $G_r(\nu)$ is the frequency dependent receiver gain as described by Eqn.~\ref{eqn:gain}, $B_r(\nu, \Omega_0-\Omega)$ is the normalised telescope beam pattern of Eqn.~\ref{eqn:BeamPattern}, $\Omega_0$ is the pointing axis of the main beam and $\Omega$ is the angular position of the satellite on the sky.

\begin{figure*}
\centering 
\includegraphics[width=0.48\textwidth]{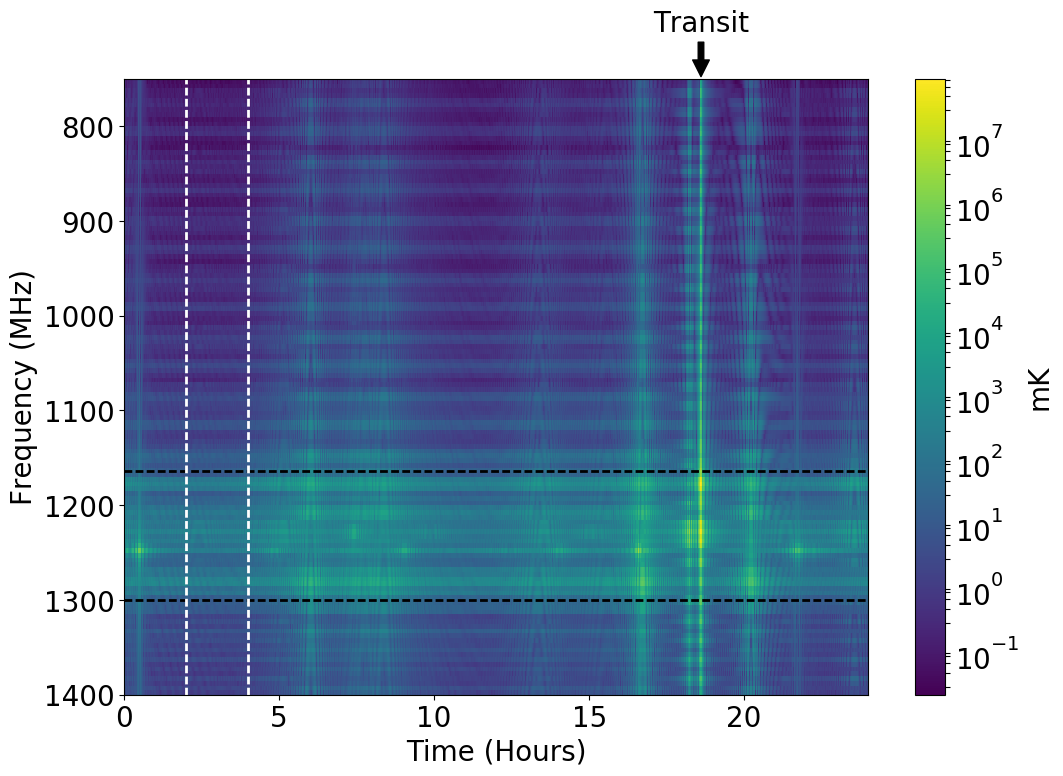}
\includegraphics[width=0.48\textwidth]{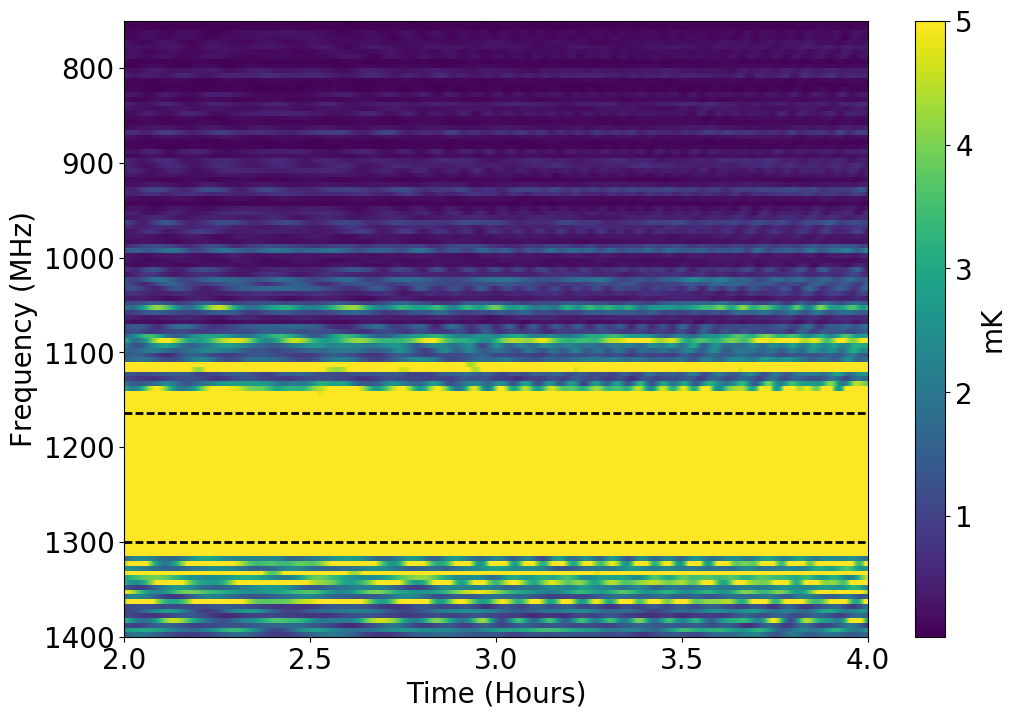}
\caption{Simulated TOD output for a drift scan observation at $-$10$^\circ$ declination, observing for a period of 24\,hours. The \textit{left} plot shows the full TOD on a log-scale. The transit event marked at the top of the \textit{left} figure indicates when a satellites is transiting the main beam of the simulated telescope. The \textit{dashed black} lines mark the frequencies 1164\,MHz and 1260\,MHz, the lower and upper frequency bounds for the L5 and L2 bands. The \textit{white dashed} lines mark a region in the TOD when no satellites approaches closer than 29$^{\circ}$ of the main beam axis. The \textit{right} plot is zoom in on the TOD between the two \textit{white dashed} lines, and uses a linear colour-scale with the regions within the GNSS band saturated to exaggerate the low level out-of-band fluctuations. }\label{fig:TOD}
\end{figure*}

The simulated contribution of all satellites at time, $t$, to the system temperature is simply the integral of Eqn.~\ref{eqn:oneSat} over all satellites above the horizon 
\begin{equation}
  T_s(t, \nu) = \sum\limits_{N_\mathrm{sats}} T_s'(\Omega, \nu),
\end{equation}
where it is assumed that $\Omega$ is a function of time and $N_\mathrm{sats}$ is approximately 30 if the horizon is at elevation 0$^{\circ}$. The \textit{left} plot of Fig.~\ref{fig:TOD} shows an example simulated TOD output for a 24\,hour drift scan observation at a fixed declination of $-$10$^{\circ}$ over a frequency range of $750 < \nu < 1400$\,MHz. At a declination of $-$10$^{\circ}$ it is expected that there will be several events where satellites pass very close to or through the main beam of the telescope. One such transit event is marked at around 18\,hours into the observation, with a peak brightness of almost 10$^{8}$\,mK when observed within the GNSS allocated frequencies. However, most important to note is that the colour-scale in the figure spans more than 8 orders-of-magnitude. During the period spanned by the two \textit{dashed white} lines in Fig.~\ref{fig:TOD} no GNSS satellites pass within 29$^\circ$ of the main beam axis. The \textit{right} plot of Fig.~\ref{fig:TOD} zooms in on this relatively quiescent region of the simulated observation. The \textit{right} plot is on a linear colour-scale and the TOD within the satellite frequency allocation is highly saturated (peaking at 800\,mK). The out-of-band regions however still show fluctuations of the order of 0.1$-$1\,mK at frequencies as low as 800\,MHz, which is still brighter than the expected HI intensity signal brightness of approximately 0.01\,mK \citep{Battye2013}. Critically, Fig.~\ref{fig:TOD} shows that the GNSS emission has a lot of spectral structure, implying it will be more challenging to remove than spectrally smooth astrophysical foregrounds. A more qualitative discussion of the relative HI and GNSS signals is given in Sec.~\ref{sec:Results}.

\ctable[
  caption = Summary of satellite output power models. Definitions of acronyms are provided in the main text and in Sec.~\ref{sec:app2} for power spectral density (PSD) definitions. Allocations are those set by ITU regulations \citep{ITU2004}.,
  label = table:power,
  width = 0.95\textwidth,
  star
]{l|l|c|c|c|c|c|r}{
	}{\FL Constellation (block) & Band     &  Allocation (MHz) & $P_t$ (dBW) & $G_t$ (dBi) & $\nu_0$ (MHz) & $f_0$ (MHz) & PSD \ML
  		GPS (IIR)               & L1 (P)   &  1563$-$1587   & 13.5 & 13.5 & 1575.42 & 1.023 & BPSK(10) \NN
                                & L1 (C/A) &  1563$-$1587   & 16.5 & 13.5 & 1575.42 & 1.023 & BPSK(1) \NN
                                & L2 (P)   &  1215$-$1239.6 & 8.75 & 13.5 & 1227.6 & 1.023 & BPSK(10) \NN
  		GPS (IIR-M)             & L1 (P)   &  1563$-$1587   & 13.5 & 13.5 & 1575.42 & 1.023 & BPSK(10) \NN
                                & L1 (C/A) &  1563$-$1587   & 16.5 & 13.5 & 1575.42 & 1.023 & BPSK(1) \NN
                                & L1 (M)   &  1563$-$1587   & 18.2 & 13.5 & 1575.42 & 1.023 & BOC(5,10) \NN
                                & L2 (P)   &  1215$-$1239.6 & 10.5 & 13.5 & 1227.6 & 1.023 & BPSK(10) \NN
                                & L2 (C/A) &  1215$-$1239.6 & 11.2 & 13.5 & 1227.6  & 1.023 & BPSK(1) \NN
                                & L2 (M)   &  1215$-$1239.6 & 15.2 & 13.5 & 1227.6  & 1.023 & BOC(5,10) \NN
  		GPS (IIF)               & L1 (P)   &  1563$-$1587   & 13.5 & 13.5 & 1575.42 & 1.023 & BPSK(10) \NN
                                & L1 (C/A) &  1563$-$1587   & 16.5 & 13.5 & 1575.42 & 1.023 & BPSK(1) \NN
                                & L1 (M)   &  1563$-$1587   & 18.2 & 13.5 & 1575.42 & 1.023 & BOC(5,10) \NN
                                & L2 (P)   &  1215$-$1239.6 & 10.5 & 13.5 & 1227.6 & 1.023 & BPSK(10) \NN
                                & L2 (C/A) &  1215$-$1239.6 & 11.2 & 13.5 & 1227.6  & 1.023 & BPSK(1) \NN
                                & L2 (M)   &  1215$-$1239.6 & 15.2 & 13.5 & 1227.6  & 1.023 & BOC(5,10) \NN
                                & L5       &  1164$-$1189   & 16 & 13.5 & 1176.45  & 1.023 & BPSK(10) \NN
  		Galileo                 & E1       &  1559$-$1591   & 15 & 15 & 1575.42 & 2.5575 & BOCc(15,2.5) \NN
								& E5A      &  1164$-$1189   & 15 & 15 & 1278.75 & 1.023 & BOCc(5,10) \NN
                                & E5B      &  1189$-$1214   & 18 & 15 & 1278.75 & 1.023 & BPSK(5) \NN
                                & E6       &  1260$-$1300   & 18  & 15 & 1191.795 & 1.023 & altBOC(10,15) \NN
		GLONASS (M)             & G1 (P)   &  1593$-$1610   & 13 & 13.5 & 1602 & 0.511 & BPSK(10) \NN
		                        & G1 (C/A) &  1593$-$1610   & 13 & 13.5 & 1602 & 0.511 & BPSK(1) \NN
		                        & G2 (P)   &  1237$-$1254   & 10 & 13.5 & 1246 & 0.511 & BPSK(10) \NN
		                        & G2 (C/A) &  1237$-$1254   & 10 & 13.5 & 1246 & 0.511 & BPSK(1) \NN
		GLONASS (K)             & G1 (P)   &  1593$-$1610   & 13 & 13.5 & 1602 & 0.511 & BPSK(10) \NN
		                        & G1 (C/A) &  1593$-$1610   & 13 & 13.5 & 1602 & 0.511 & BPSK(1) \NN
		                        & G2 (P)   &  1237$-$1254   & 10 & 13.5 & 1246 & 0.511 & BPSK(10) \NN
		                        & G2 (C/A) &  1237$-$1254   & 10 & 13.5 & 1246 & 0.511 & BPSK(1) \NN
      			                & G3 (P)   &  1189$-$1214   & 14.8 & 13.5 & 1207.14 & 1.023 & BPSK(10) \LL}
		
There are several simplifications in these GNSS transmission models that are neglected. First, no atmospheric absorption is assumed as that is expected to induce only a 0.04\,dB of loss at L-band frequencies \citep[e.g.][]{BigotSazy2015,Steigenberger2017}. Variations due to the beam pattern of the satellites are also neglected as the beam is designed as such that there is no more than a factor of two difference between transmissions at the zenith and the horizon \citep{Montesano2007,Marquis2015}. There is also a GPS L3 service at 1381.05\,MHz that is not modelled as this is an intermittent broadcast. The variance in the power output levels of satellites within constellations, or blocks, are of the order of 10\,per\,cent for GPS and Galileo and up to a factor of 2 for GLONASS broadcasts as presented in \citet{Steigenberger2017}, these variations are not included. Finally, the GLONASS G3 signals are simulated at only one central frequency. In reality the GLONASS G3 (and G1) services avoid self interference by transmitting at carrier frequencies unique to each satellite within the allocated ITU band\citep{hofmann2007}. In these simulations we have assumed one nominal central frequency, which is why the peak power within the G3 band is slightly offset from the band centre in Fig.~\ref{fig:Bright1}.

\section{Results}\label{sec:Results}

The most obvious way to mitigate RFI from GNSS satellites when observing with a single-dish telescope is to simply excise, or \textit{flag}, TOD when a satellite is transitting through or close to the main beam of the telescope. However, these results will show that the fluctuations of the emission from satellites moving within the far sidelobes, more than 20$^\circ$ away from the main beam axis, may still be problematic when integrated over the long periods required to detect the HI intensity signal. Sec.~\ref{sec:drift} will discuss the impact of the GNSS satellites at the TOD level using simulated drift scan observations with an SKA dish; and Sec.~\ref{sec:survey} will show the simulated impact of the integrated GNSS emission on a single-dish SKA HI IM survey covering 30000\,sq.\,deg. with 200\,dishes. 

Note that the results presented in this section excise satellites approaching close to the main beam by modifying the main beam response as discussed in Sec.~\ref{sec:beam}. All results presented in this Section use the filtered beam responses out to radii of $\pm$5$^\circ$, 10$^\circ$, 15$^\circ$ and 20$^\circ$ from the main beam.

\subsection{SKA Drift Simulation}\label{sec:drift}

\begin{figure}
\centering 
\includegraphics[width=0.5\textwidth]{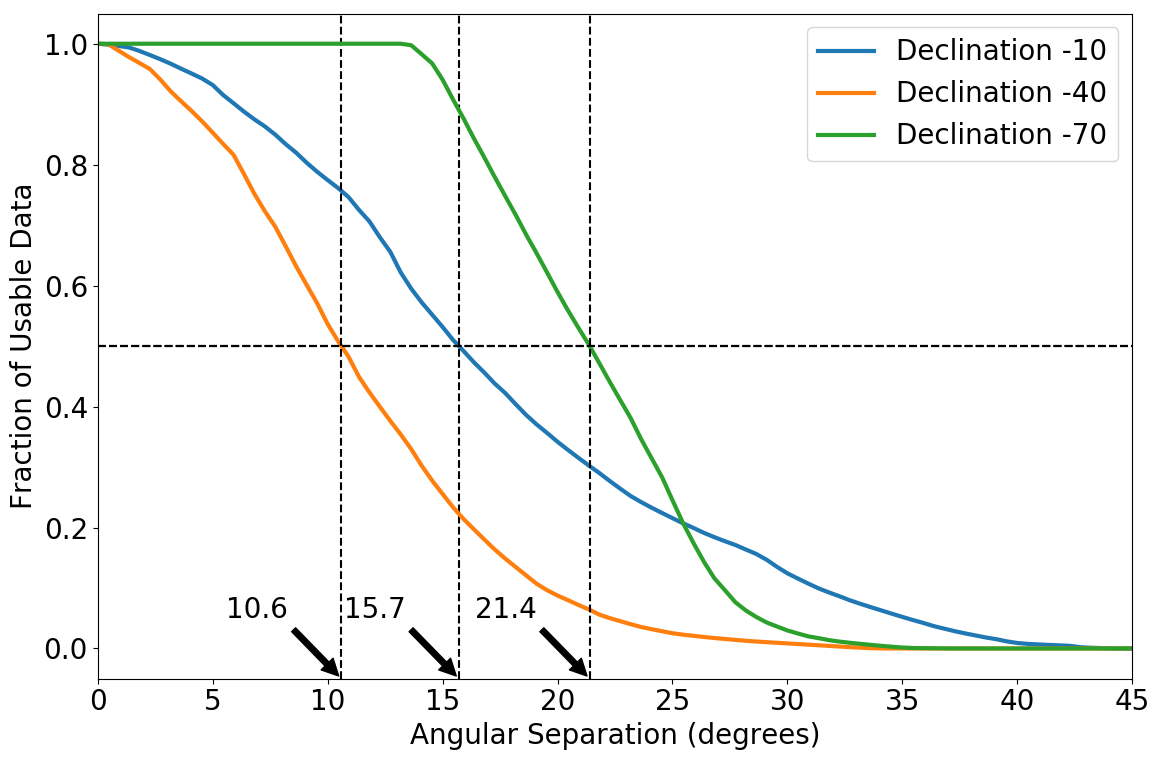}
\caption{Fraction of usable data remaining after excising all satellites within a given angular separation of the main beam. For the SKA less than 2\,per\,cent of data will contain a satellite transiting the main beam. }\label{fig:ResultsFlag}
\end{figure}

In this section we will simulate drift scan observations, where the telescope is fixed in the horizon coordinate frame and a circle on the sky is mapped at a fixed declination using the sidereal motion of the Earth. Three drift scans observations were simulated such that two lay within the declination range of the satellites at $-$10$^{\circ}$ and $-$40$^{\circ}$, and one below the minimum GNSS declination at $-$70$^{\circ}$. Fig.~\ref{fig:ResultsFlag} shows how much data would be lost when excising satellites within some angular separation from the main beam axis for the three different drift scan observations. The main beam in these simulations has an approximate full-width half-maximum (FWHM) of 1$^\circ$, and Fig.~\ref{fig:ResultsFlag} shows that for all three drift scans very little data is lost to main beam transits (approximately 2\,per\,cent at $-$40$^{\circ}$ declination). However, GNSS transmissions are sufficiently bright (see Fig.~\ref{fig:Bright1}), that it is likely necessary to excise data containing satellite transits far further from the main beam axis. Fig.~\ref{fig:ResultsFlag} shows the angular separation at each declination that would result in 50\,per\,cent loss of data, which lies between 10$^{\circ}$$-$20$^{\circ}$ depending on declination. The fractional loss of data when using an observing strategy that spans multiple declinations would be a weighted sum of the curves shown in Fig.~\ref{fig:ResultsFlag}.

\begin{figure*}
\centering 
\includegraphics[width=0.95\textwidth]{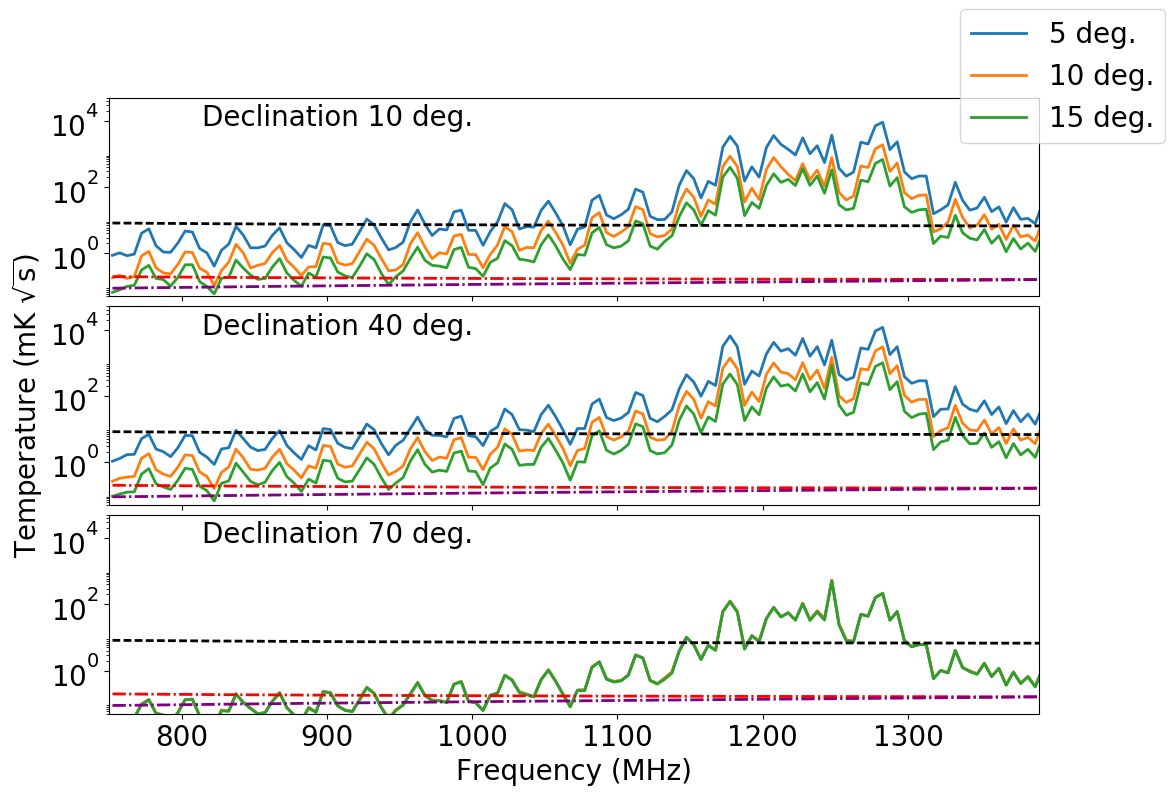}
\caption{The expected RMS distribution of the simulated GNSS signals in the TOD for several fixed declination drift scan observations over one sidereal day. The three coloured lines shows the predicted RMS assuming all satellites within $\pm$5, 10 and 15$^\circ$ of the main beam are excised from the data. The \textit{dashed black} line is the expected SKA sensitivity, and the \textit{dot-dashed red} line is the expected survey sensitivity assuming 30\,days observing with 200\,dishes over 30000\,sq.\,deg.}\label{fig:ResultsFrequencies}
\end{figure*}

One important metric to determine is the scale of the fluctuations of the GNSS satellite transmissions when picked up within the sidelobes of the telescope relative to both the expected RMS of the HI intensity signal and the noise of an SKA receiver. To calculate the approximate noise level of the SKA receiver we adopt the system temperature model described in \citet{Bull2015}
\begin{equation}
		T_\mathrm{sys} = T_\mathrm{rx} + T_\mathrm{sky},
\end{equation}
where $T_\mathrm{rx} = 20$\,K and $T_\mathrm{sky} = 60\text{\,K} (\nu / 300\text{\,MHz})^{-2.5}$. The subsequent RMS of the thermal noise fluctuations is then calculated using the well known radiometer equation
\begin{equation}\label{eqn:sigmaw}
	\sigma_{w} = \frac{T_\mathrm{sys}}{\sqrt{\delta \nu }} 
\end{equation}
where $\delta\nu$ is the channel width given in Table~\ref{table:survey}. It is also informative to compare the scale of the GNSS fluctuations to the final RMS of a nominal SKA HI IM survey that observes 30000\,sq.\,deg., with 200\,dishes over a period of 30\,days \citep{Santos2015,Bull2015,Harper2017}. The survey RMS is estimated as 
\begin{equation}
	\sigma_\mathrm{surv} = \sigma_w \sqrt{\frac{\Omega_\mathrm{surv}}{N_d T_\mathrm{surv} \Omega_\mathrm{beam}  } } ,
\end{equation}
where $\sigma_w$ is as defined by Eqn.~\ref{eqn:sigmaw}, $\Omega_\mathrm{surv}$ is the survey area, $\Omega_\mathrm{beam}$ is the beam area (assuming here a FWHM of 1$^\circ$), $N_d$ is the number of dishes in the survey and $T_\mathrm{surv}$ is the survey observing time. To calculate the expected RMS of the HI signal we adopt the HI model described in \citet{Harper2017} and integrate over the angular power spectrum  at each frequency as
\begin{equation}
	\sigma^2_\mathrm{HI}(\nu) = \frac{1}{4 \pi}\sum\limits_{\ell=0}^{\infty} (2\ell + 1) C_\ell(\nu) ,
\end{equation}
where $C_\ell(\nu)$ is the simulated HI model angular power spectrum at frequency $\nu$ and $\ell$ is the spherical harmonic coefficient.

Fig.~\ref{fig:ResultsFrequencies} shows the mean RMS for the drift scan observations when suppressing satellites that transit within 5, 10 or 15$^\circ$ of the main beam axis. The figure shows that declinations of $-$10$^{\circ}$ and $-$40$^\circ$ the satellite fluctuations only exceed the thermal noise of a single SKA receiver (the \textit{black dashed} line) at the edge of the lower GNSS bands around frequencies of $\nu > 1150$\,MHz. The \textit{red dot-dashed} line in the figure shows for comparison the expected SKA HI IM survey noise level, which is less than the expected RMS of the GNSS emission at nearly all frequencies, even when observing near the South celestial pole (SCP) at $-70^{\circ}$ declination. This is also the case for the RMS of the HI signal, which, as should be expected, is comparable to the survey noise RMS.

\begin{figure}
\centering 
\includegraphics[width=0.5\textwidth]{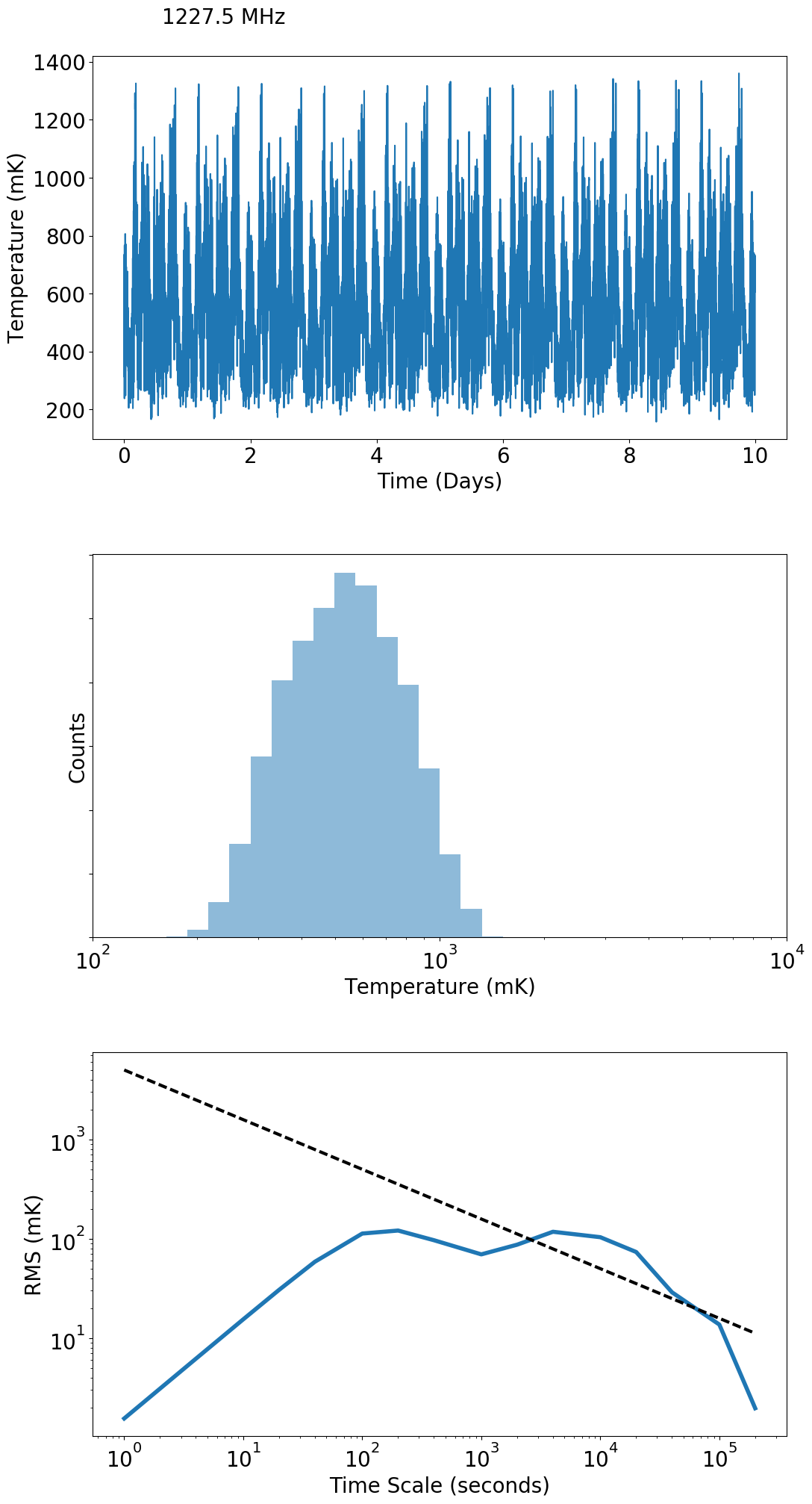}
\caption{Statistical description of one simulated 5\,MHz channel centred at 1227.5\,MHz. The \textit{top} plot shows the timeline of the received GNSS fluctuations. The \textit{middle} plot shows the log-normal distribution of the samples in the \textit{top} plot. The \textit{bottom} figure shows the Allan deviation. This shows how the RMS of the fluctuations decreases with integration time is similar to white noise on long time scales (the \textit{dashed} line).}\label{fig:ResultsHistogram}
\end{figure}

Fig.~\ref{fig:ResultsHistogram} shows an example TOD stream of the simulated GNSS fluctuations alongside a histogram of the fluctuations and the Allan variance of the TOD using the beam model filtered out to 15$^\circ$ from the main beam. The \textit{top} plot shows the variations in the TOD in the 1227.5\,MHz frequency channel over a ten day period. Note that there is clearly a periodicity to the signal, which corresponds to approximately 12\,hour orbital periods of the GNSS satellites. The next plot shows that the distribution of the GNSS signal is log-normal. The last plot shows the Allan RMS, and it is here that the statistical properties of GNSS emissions is of concern. The Allan RMS is defined as the RMS measured on different timescales within the data, and for Gaussian white noise the Allan RMS should drop as $1/\sqrt{T}$ as illustrated in the figure by the \textit{black dashed} line. However, the measured RMS of the satellites is seen to increase up to timescales of approximately 100\,seconds, which is representative of the transit time of the satellite through the sidelobes of the telescope. Then on intermediate timescales, between 100\,seconds and 12\,hours, the RMS appears largely flat and then on timescales over 12\,hours the RMS falls off. This figure implies therefore that only on integration times of a day or more will the \textit{pseudo}-random variations of the GNSS satellites integrate down like white noise.

\begin{figure}
\centering 
\includegraphics[width=0.5\textwidth]{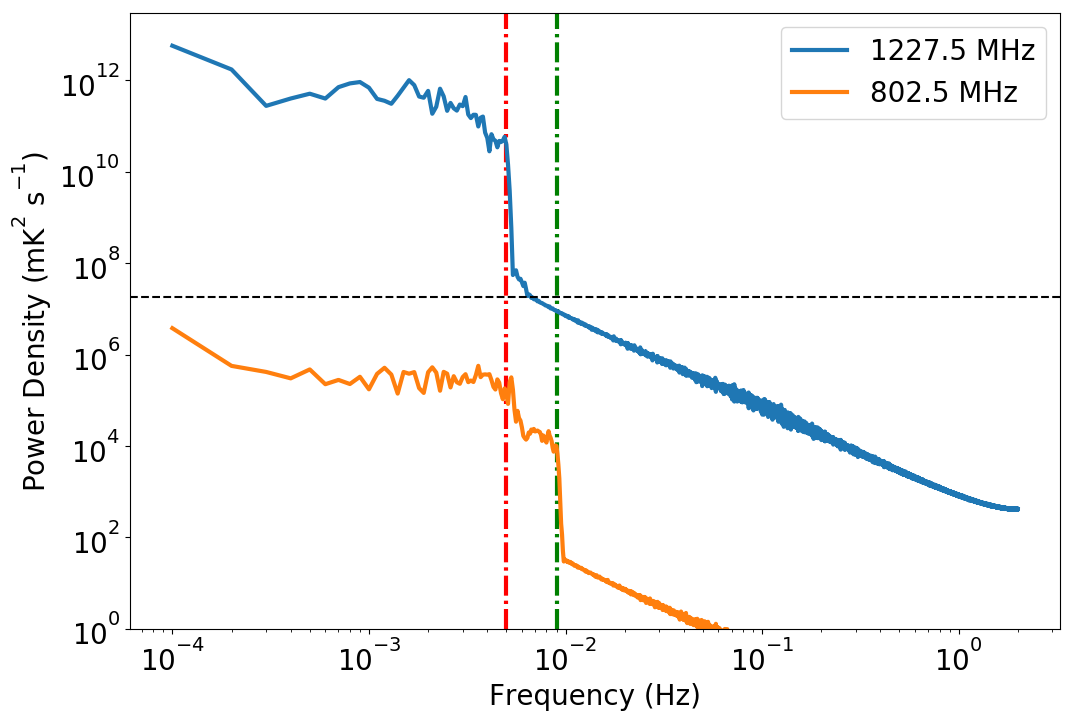}
\caption{Power spectral densities of the simulated GNSS fluctuations from the declination $-$70$^\circ$ drift scan. The top power spectrum is of the 1227.5\,MHz channel, and the bottom is the 802.5\,MHz channel. The \textit{black dashed} line shows the expected mean RMS level across the whole simulated bandpass. The \textit{red} and \textit{green dot-dashed} lines mark the approximate timescale for the satellites to transit the beam sidelobes. As the beam is smooth on times scales shorter than the transit time there is no power in the fluctuations in the GNSS emission.}\label{fig:ResultsPowerSpec}
\end{figure}

Complementary information to the Allan RMS in Fig.~\ref{fig:ResultsHistogram} can be drawn from taking the power spectrum of the GNSS fluctuations as shown in Fig.~\ref{fig:ResultsPowerSpec}. The figure shows the mean power spectra for channels 1227.5 and 802.5\,MHz, taken from the drift scan centred on $-$70$^\circ$ declination for a period of 10000\,seconds. The first feature, marked in the figure by the \textit{red} and \textit{green dot-dashed} lines, are the sharp cut-offs in power on short timescales (approximately $< 100$\,seconds). This sharp cut-off is due the convolution of the sidelobes in the beam antenna pattern  with the fluctuations in the GNSS power at a given time. The two frequencies have different cut-offs because the sidelobe structure is changing in frequency, for example the higher frequencies have narrower sidelobes and thus a shorter satellite transitting time. The other key feature of Fig.~\ref{fig:ResultsPowerSpec} is the leveling off in the power on long timescales, which appear to be relatively flat (e.g., white noise) except for a small turn up on the very largest timescale. Combining the Allan RMS of Fig.~\ref{fig:ResultsHistogram} with the power spectra just discussed implies, reassuringly, that on timescales longer than a few minutes the GNSS transmissions should approximately integrate down as Gaussian white noise in time.

\subsection{SKA Survey Simulation}\label{sec:survey}

\begin{figure*}
\centering 
\includegraphics[width=0.85\textwidth]{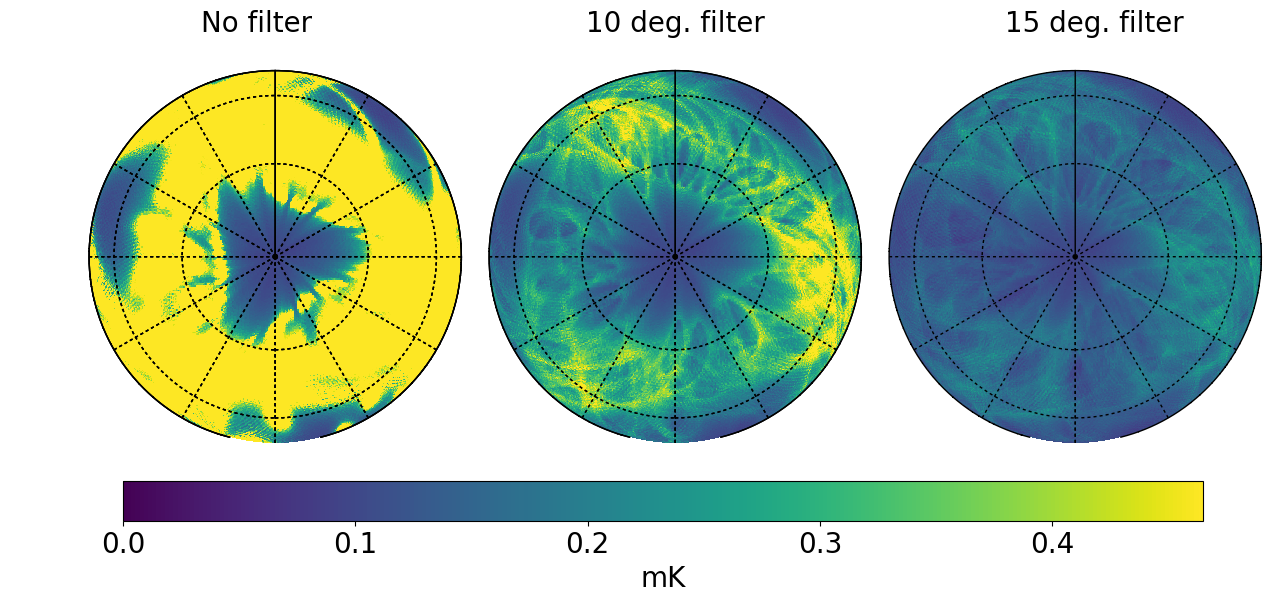}
\includegraphics[width=0.85\textwidth]{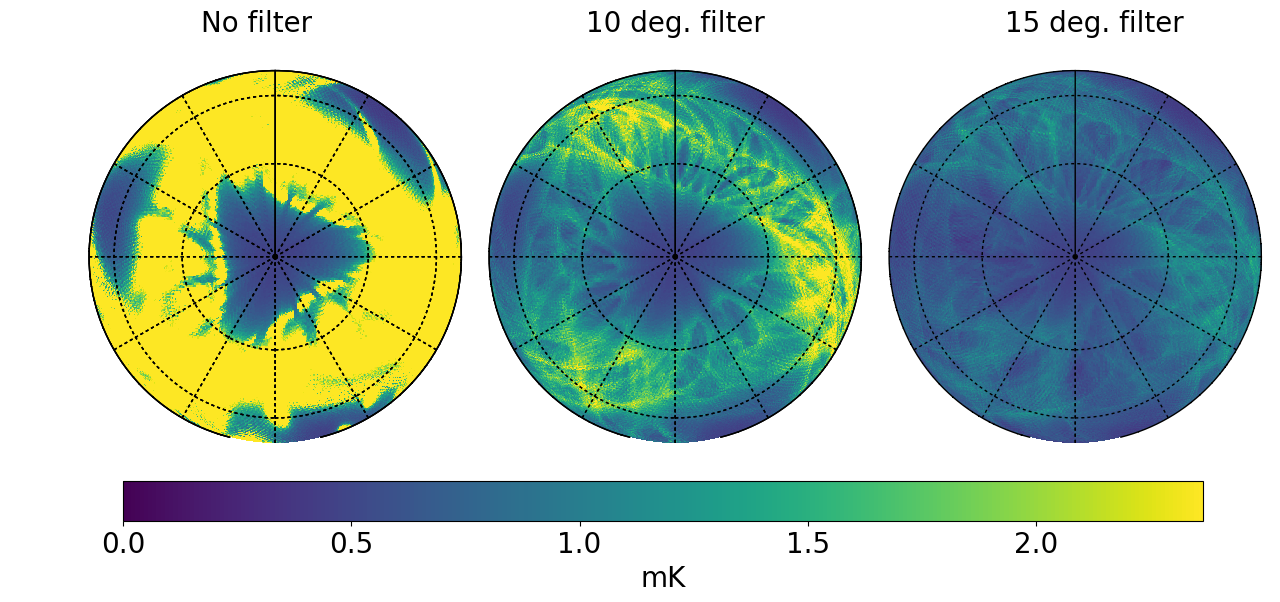}
\includegraphics[width=0.85\textwidth]{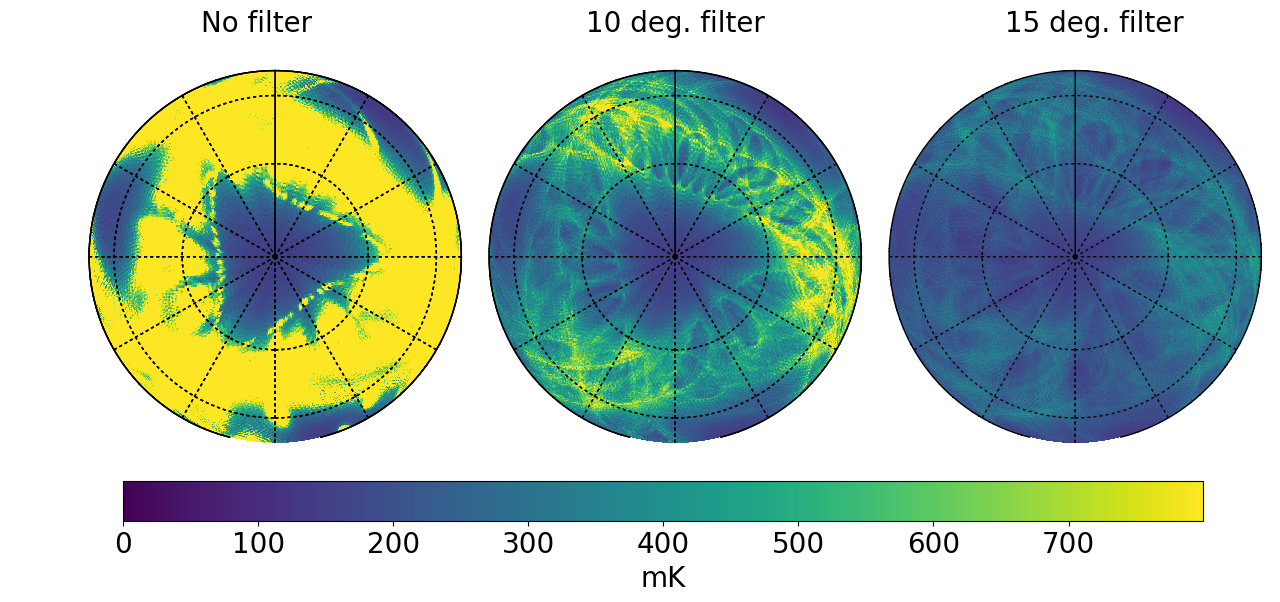}
\caption{Orthognomic projections of the expected integrated GNSS sky emission for a single telescope observing for 90\,days centred on the SCP and increasing in declination radially to the celestial equator at the edge. From \textit{top} to \textit{bottom} the central frequencies of the maps are 750, 1000 and 1200\,MHz. From \textit{left} to \textit{right} different beam filters are applied to the data from no filter, a 10$^\circ$ filter and a 15$^\circ$ filter. The maps are all normalised relative to the RMS of the 10$^\circ$ filtered plots at each frequency. The bandwidth of each map is 5\,MHz. Due to the high-dynamic range of the data all images use a log colour scale.}\label{fig:Results1}
\end{figure*}

Now we want to determine how the observed celestial sky will appear for just the integrated emission from the GNSS satellites. This is done by integrating the survey observing strategy described in Section~\ref{sec:surveydesign}. Fig.~\ref{fig:Results1} shows the integrated Southern celestial hemisphere, in an orthognomic projection, where the centre of each image is the SCP and the declination increases radially outwards to the celestial equator. The columns in Fig.~\ref{fig:Results1} show the impact of filtering the satellites on the integrated GNSS emission for no filter, a 10$^{\circ}$ filter and a 15$^{\circ}$ filter. Each row has a different central frequency, with the \textit{top} two rows centred at frequencies considered \textit{out-of-band} for GNSS emission. The central frequencies are 750, 1000 and 1200\,MHz respectively, each using a 5\,MHz channel width.

One key feature of Fig.~\ref{fig:Results1} is the difference in scale between the two out-of-band GNSS frequency channels are a factor of two or three different in brightness over 250\,MHz. However, between 1000 and 1200\,MHz (e.g., moving from out-of-band to in-band) there is a two order-of-magnitude difference in the expected brightness. The spatial distribution of the satellites changes very little with frequency, and as can be inferred from Fig.~\ref{fig:Orbits1}, there are several relatively clear regions of the sky, one around the SCP, and several others positioned equidistance apart centred on $\approx -30^{\circ}$ declination. As these spatial structures are dependent on the orbital parameters of the satellites they will not change when using a different observing strategy. The sky area not contaminated by GNSS satellites around the SCP (declination < $-65^{\circ}$) is approximately 1900\,sq.\,deg., which is less than a tenth of the nominal SKA HI IM survey sky area, and a fifth of the sky area proposed for the precursor MeerKAT survey \citep{Santos2017}. Even though this sky area is far smaller when compared to the currently proposed SKA HI IM survey areas, the significant reduction in GNSS contamination may make this a desirable survey field.

\begin{figure}
\centering 
\includegraphics[width=0.5\textwidth]{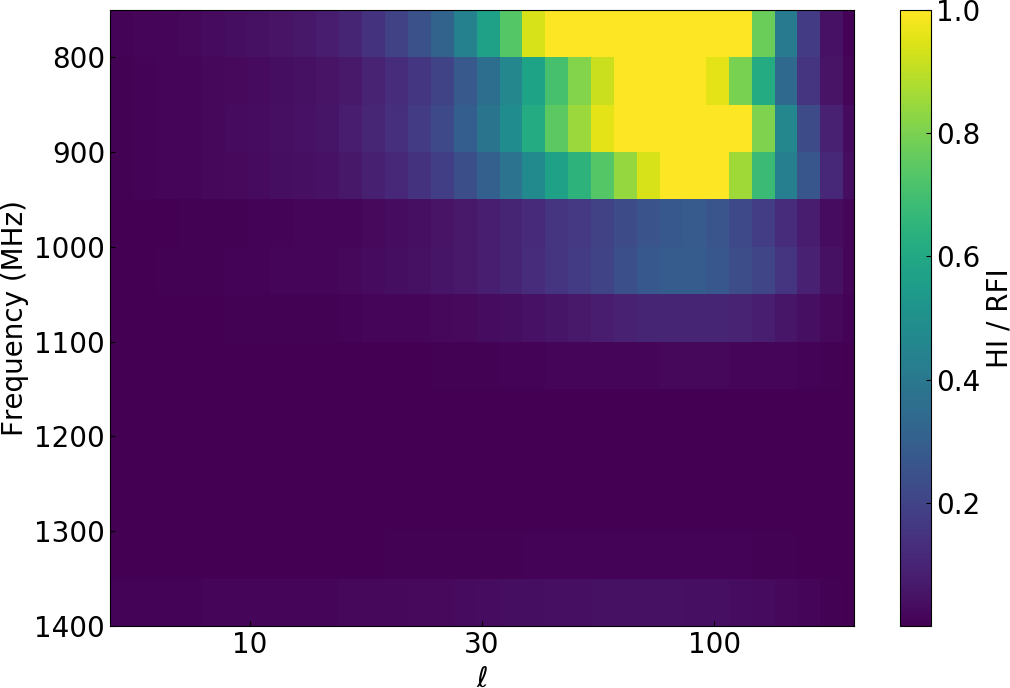}
\caption{Ratio of the expected HI angular power spectrum at each simulated frequency over the measured GNSS fluctuation angular power spectrum. The figure is saturated at a ratio of unity, with dark regions representing scales and frequencies that are likely to be unusable without careful consideration and removal of the GNSS signals. }\label{fig:Results2}
\end{figure}

Fig.~\ref{fig:Results1} shows that there will be a large-scale feature across the sky that is caused by the GNSS satellite emission. As the structure of the GNSS signal is dictated by the orbital paths of the satellites, it will also not average down with time. To evaluate the impact of this large-scale feature on a future SKA HI IM survey it is useful to compare the angular power spectra of the GNSS emission at each frequency to the angular power spectra of the expected HI signal. The angular power spectra of the GNSS satellite emission is measured by performing a spherical harmonic transform on the maps shown in Fig.~\ref{fig:Results1} \citep[e.g.,][]{Peebles1973}
\begin{equation}\label{eqn:GNSSCl}
	G_\ell = \frac{f_\mathrm{sky}}{2 \ell + 1}\sum\limits_{m} \left| a_{\ell m} \right|^2,
\end{equation}
where $G_\ell$ is the angular power spectrum, $a_{\ell m}$ are the spherical harmonic coefficients, and a correction factor of $f_\mathrm{sky}$ is included to account for only a partial sky being observed. The \texttt{PolSpice} package \citep{Chon2004} is used to perform the spherical harmonic transforms of the GNSS emission for each frequency and beam filter.

To calculate the HI angular power spectrum we adopt the simple model for the mean HI brightness ($\bar{T}_\mathrm{obs}$) given by \citep{Battye2013}
\begin{equation}
\bar{T}_{\mathrm{obs}}(z) = 44\,\mu\mathrm{K} \left(\frac{\Omega_{\mathrm{HI}}(z)h }{2.45 \times 10^{-4}} \right) \frac{(1 + z)^2}{E(z)} ,
\end{equation}
where $\Omega_\mathrm{HI}$ is the neutral HI fraction at redshift $z$ and $E(z)$ describes the Hubble expansion. For these simulations no evolution of the neutral HI is modelled, and a constant value of $\Omega_\mathrm{HI} = 6.2 \times 10^{-4}$ is assumed \citep{Switzer2013}. The angular power spectra of the HI at each frequency is then modelled as 
\begin{equation}\label{eqn:HICl}
C_{\ell} = \frac{H_0}{c} \int \mathrm{d}z E(z) \left[ \frac{\bar{T}_{\mathrm{obs}}(z) D(z)}{r(z)} \right]^2 P_{\mathrm{cdm}}\left(\frac{\ell + 0.5}{r} \right)
\end{equation}
where $P_\mathrm{cdm}$ is the underlying cold dark matter transfer function generated using \texttt{CAMB} \citep{Lewis2002}, $r(z)$ is the comoving distance, and $D(z)$ is the growth factor. The Limber approximation of $k \approx \frac{\ell + 0.5}{r}$  \citep{Datta2007} is also assumed in Eqn.~\ref{eqn:HICl}.

The figure-of-merit used to determine the impact of the GNSS emission on a possible SKA HI IM survey is simply the ratios of the angular power spectra described by Eqn.~\ref{eqn:GNSSCl} and Eqn.~\ref{eqn:HICl}. The ratio is calculated at each frequency as
\begin{equation}\label{eqn:ResultsAngularSpectra1}
	M_\ell(\nu) = \frac{C_\ell(\nu) B_\ell(\nu)}{G_\ell(\nu)},
\end{equation}
where the HI spectrum is multiplied by an additional factor of $B_\ell$ to account for the angular power of the SKA primary beam. Fig.~\ref{fig:Results2} shows how Eqn.~\ref{eqn:ResultsAngularSpectra1} varies with angular harmonic $\ell$ and frequency $\nu$ when filtering GNSS satellites that pass within 15$^{\circ}$ of the main beam axis. The figure shows that the power in the GNSS emission exceeds the expected HI power on all scales observable with the beam of a single SKA telescope for SKA band 2 frequencies ($950 < \nu < 1400$\,MHz). Even at the lower band 1 frequencies shown in the plot ($750 < \nu < 950$\,MHz) the GNSS emission angular power exceeds the HI power on the largest scales of $\ell < 30$. Fig.~\ref{fig:Results2} should therefore serve as a warning that careful consideration should be given to developing methods for suppressing GNSS emission within HI IM observations.

%
%

\section{Conclusions and Discussion}\label{sec:conclusion}


This paper has presented simulations of GNSS satellites within the GPS, GLONASS and Galileo constellations. Simulated drift scan and survey observations were used to demonstrate the statistical properties of the satellite transmissions within the sidelobes of a model SKA beam. It was shown in Fig.~\ref{fig:ResultsFlag} that transits of GNSS satellites through main beam will not exceed more than 2\,per\,cent of the data. In Sec.~\ref{sec:survey} it was demonstrated that the emission GNSS satellites will imprint large-scale structures onto sky when performing a large-area HI IM survey. These structures are not homogeneous or Gaussian in distribution, and as such some regions of the sky are more contaminated than others, with a large 1900\,sq.\,deg region around the SCP identified as the cleanest region. These  large-scale structures due to integrated GNSS emission is not expected to average down with time, though will become smoothed on small-scales. Finally, in Fig.~\ref{fig:Results2} it was shown that the expected angular power of the integrated GNSS emission will exceed the expected HI angular power on all scales sensitive to the SKA for frequencies $>950$\,MHz, and is still problematic on larger scales at lower frequencies. This implies that GNSS emission, even when excised from the data up to 20$^\circ$ from the main beam will be problematic for future HI IM surveys.

One method for removing the integrated emission from GNSS satellites may be to apply existing component separation methods used to suppress astrophysical foreground contaminants, such as generalised needlet internal linear combination \citep{Olivari2016} and others. However, many of these component separation methods rely on astrophysical foregrounds being spectrally smooth, but as shown in Fig.~\ref{fig:Bright1}, Fig.~\ref{fig:TOD} and Fig.~\ref{fig:ResultsFrequencies} the spectral structure of GNSS emissions is far from spectrally smooth. The effectiveness of existing component seperation methods may be limited without some modifications to include prior information of the GNSS emissions spectral structure.

It is possible to design hardware solutions to mitigate the impact of GNSS emission. It has been demonstrated in the past that cross-correlating data from auxiliary telescopes that are tracking GNSS satellites \citep{Galt1991}, or with hardware simulated GNSS signals \citep{ellingson2001} with data from the primary observing dish can significantly suppress GNSS interference. It has also been shown that phased array feeds (PAFs) can perform spatial filtering to adaptively suppress transmissions from GNSS satellites, which has been demonstrated with the Australian SKA Pathfinder (ASKAP) \citep{hellbourg2012,hellbourg2014}. GNSS emission can also be suppressed by building a bespoke HI IM experiment and designing in strict requirements on beam sidelobe suppression such as with the BINGO telescope \citep{Battye2013}.

Finally, in the real observations there will be many complexities that may modify the results presented in Sec.~\ref{sec:Results}. For example, we have modelled the SKA beam using a simple Gaussian tapered airy disk that is the same for all telescopes. In reality the beam will have more complex sidelobe structures \citep{davidson2013} and each telescope beam will have small differences. This will change the way the spatial structures average on the sky perhaps making them more complex but could also be beneficial by reducing the overall angular power of the GNSS emission fluctuations. There are also many other complications that we do not consider here, such as polarisation leakage,  satellite beams, orbital perturbations, power variations between satellites and more, all of which may make the GNSS interference more complex.

\section*{Acknowledgements}

SH and CD acknowledge support from an STFC Consolidated Grant (ST/P000649/1). SH and CD also acknowledge support from an ERC Starting (Consolidator) Grant (no.$\sim$307209) under FP7. The authors would also like to thank Richard Battye and Ian Browne for helpful discussions during the writing of this manuscript. We would also like to thank Lucas Olivari for contributing the HI simulations. We would like to thank Mario Santos, Khan Asad and Oleg Smirnov for providing the MeerKAT beam measurements. Finally, we would like to thank the useful comments received from our reviewer.

\appendix
\section{GNSS Orbits}\label{sec:app1}

The orbit of each satellite at any given time can be calculated from periodically measured variables contained within a data structure known as the two-line element (TLE). The five key parameters describing the orbital plane of a satellite in a TLE are:
\begin{itemize}
	\item $P$ - Mean motion in revolutions per day.
    \item $M_0$ - Mean anomaly at the time of TLE.
    \item $e$ - Eccentricity of the orbit.
    \item $i$ - Orbital inclination.
    \item $\Omega$ - Longitude of the Ascending Node (relative to a right ascension of zero).
\end{itemize}
Each TLE also contains a time of measurement and the argument of perigee, which together allow for the location of the satellite to propagate along its orbital path. The coordinate transfrom described in this section is based on the description given in \citet{green1985}. The TLEs for each the satellites in each GNSS constellation was taken from the online repository \url{www.celestrak.com}.

The first step is to determine the angular position of the satellite in its orbit, referred to as its \textit{true anomaly}, at time $t_i$ relative to some start time $t_0$. This is derived from the \textit{mean anomaly} at time $t_0$ using the know mean motion of the satellite 
\begin{equation}
	M = M_0 + 2 \pi P  (t_i - t_0) \,\mathrm{mod}\, 2 \pi .
\end{equation}
The mean anomaly at time $t_i$ is related to the eccentric anomaly by
\begin{equation}\label{eqn:EccentricAnomaly}
	M = E - e \sin(E) ,
\end{equation}
where $e$ is the orbital eccentricy. Solving Eqn.~\ref{eqn:EccentricAnomaly} for $E$ is done iteratively using the Newton-Raphson method. The \textit{true anomaly} ($\nu$) is then found using  
\begin{equation}
	\nu = 2 \tan^{-1}\left( \sqrt{ \frac{1 + e}{1 - e} } \right) \tan(0.5 E)
\end{equation}
which relates then to the position angle of the satellite in its orbit at time $t_i$ by $\theta = \nu + \omega$, where $\omega$ is the argument of perigee. 

To get the cartesian position vector of the satellite at position angle $\theta$ first the semi-major axis ($a$) must be calculated using the mean motion of the satellite ($P$) and Kepler's second law
\begin{equation}
	a = \left(\frac{\mu}{4 \pi^2 P^2} \right)^{1/3} ,
\end{equation}
where $\mu$ is the combination of the Gravitational constant $G$ and the Earth's mass $M_E$.  The instantaneous radial distance of the satellite from the centre of the Earth is
\begin{equation}
	r = a \left(1 - e \cos(E)\right),
\end{equation}
where $e$ and $E$ are eccentricy and the eccentric anomaly respectively as before. Finally, the position vectors are found via three transformations:
\begin{equation}
 x = r \left( \cos(\nu + \omega) \cos(\Omega)  - \sin(\nu + \omega) \sin(\Omega) \cos(i) \right), 
\end{equation}
\begin{equation}
 y = r \left( \cos(\nu + \omega) \sin(\Omega)  - \sin(\nu + \omega) \cos(\Omega) \cos(i) \right),
\end{equation}
\begin{equation}
 z = r \sin(\nu + \omega) \sin(i),
\end{equation}
where all parameters are as previously defined and $i$ is in the orbital inclination.

Next, to estimate the position of the satellites in the horizon frame of the telescope the position of the observer in the non-rotating Earth coordinate frame must be calculated. The position of the telescope at an arbitrary altitude is defined as
\begin{equation}
 x_t = (R_t + f R_E) \cos(\theta + \mathrm{GST}) \cos(\phi),
\end{equation}
\begin{equation}
 y_t  = (R_t + f R_E) \sin(\theta + \mathrm{GST}) \cos(\phi),
\end{equation}
\begin{equation}
 z_t  = (R_t + f R_E) \sin(\phi),
\end{equation}
where $R_t$ is the altitude of the observatory, $R_E$ is the radius of the Earth, $\phi$ is the observer longitude, and $\theta$ is the observer latitude. To fix the coordinate system to the celestial frame the system is rotated in longitude using the he Greenwich mean sidereal time (GST). The factor $f$ accounts for the flattening of the Earth at the poles and is defined by the World Geodetic System 1984 (WGS-84) as
\begin{equation}
	f = \frac{1}{\sqrt{ 1 + (1/F - 2) \sin^2(\phi)/F }}
\end{equation}
where $F \approx 298.25$. From the position vectors of the satellites and the observer at time $t_i$ it is trivial to calculate the satellites line-of-sight distance, azimuth and elevation positions.

\section{GNSS Power Spectral Density Models}\label{sec:app2}

In this section we will define the PSD models used for each satellite service as described in Table~\ref{table:power}. As discussed in Sec.~\ref{sec:bright} the most basic PSD is that of satellites using simple binary phase-shift keying (BPSK) modulation, which has a PSD defined by a sinc$^2$ function
\begin{equation}\label{eqn:app1}
	S(\nu, m)_\mathrm{BPSK} = \frac{\mathrm{sinc}^2(\pi \left[\nu - \nu_c \right] / m \nu_0)}{ m \nu_0 } ,
\end{equation}
where $\nu_c$ is the carrier frequency, which is some integer multiple of $\nu_0$. The \textit{chip rate}, which defines the phase-switching frequency, is defined as $m \nu_0$. 

For services provided by Galileo and the GPS military (M) code the BPSK signals are modulated by a rectangular subcarrier wave, this is referred to as binary offset carrier (BOC) modulation. The purpose of BOC modulation is to spread the peak PSD frequencies away from $\nu_c$. For these simulations we will define three forms of BOC modulation: standard BOC, cosine-BOC (BOCc) and alternative BOC (altBOC). Details of the differences in these modulation schemes can be found in \citet{hofmann2007}, here we will simply describe the PSD models. 

\begin{figure}
\centering 
\includegraphics[width=0.5\textwidth]{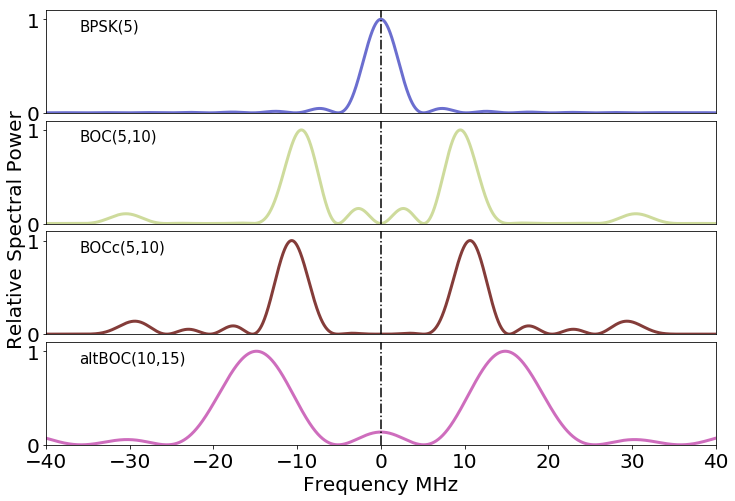}
\caption{The four PSD models used to simulate the total power emission from the GNSS satellites. The spectra are normalised to unity at the peak and are all shown on the same linear scale. The four spectra are representative of the Galileo E5A (BOCc(5,10)), E5B (BPSK(5)), E6 (altBOC(10,15)), and GPS-M (BOC(5,10)) spectra. The \textit{dashed} black line marks the zero frequency.}\label{fig:app2fig1}
\end{figure}

The standard BOC modulation PSD is defined as \citep{hofmann2007}
\begin{equation}
	S(\nu, m, n)_\mathrm{BOC} = m \nu_0  \left( \frac{\sin(\pi \frac{\left[\nu - \nu_c \right]}{2  n  \nu_0}) 
	\sin(\pi \frac{\left[\nu - \nu_c \right]}{m  \nu_0})}
	{\pi \nu \cos(\pi \frac{\left[\nu - \nu_c \right]}{2  n  \nu_0})} \right)^2,
\end{equation}
where as in Eqn.~\ref{eqn:app1} $\nu_c$ is the carrier frequency, which is an integer multiple of the fundamental frequency $\nu_0$. The chipping rate of the underlying BPSK signal is still $m \nu_0$, while the subcarrier frequency is $n \nu_0$. The BOCc PSD is defined as
\begin{equation}
	S(\nu, m, n)_\mathrm{BOCc} = m \nu_0  \left( \frac{\sin^2(\pi \frac{\left[\nu - \nu_c \right]}{4  n  \nu_0}) 
	\cos(\pi \frac{\left[\nu - \nu_c \right]}{m  \nu_0})}
	{\pi \nu \cos(\pi \frac{\left[\nu - \nu_c \right]}{2  n  \nu_0})} \right)^2 , 
\end{equation}
and the alternative BOC (altBOC) modulation with the spectrum described as
\begin{equation}
	S(\nu, m, n)_\mathrm{altBOC} = m \nu_0  \left( \frac{\sin^2(\pi \frac{\left[\nu - \nu_c \right]}{m  \nu_0}) 
	 \left[ 1 - \cos(\pi \frac{\left[\nu - \nu_c \right]}{2 n  \nu_0})\right]}
	{\pi \nu \cos(\pi \frac{\left[\nu - \nu_c \right]}{2  n  \nu_0})} \right)^2 .
\end{equation}
Fig.~\ref{fig:app2fig1} shows a comparison between these four PSD using the Galileo E5A (BOCc(5,10)), E5B (BPSK(5)), E6 (altBOC(10,15)), and GPS-M (BOC(5,10)) codes as examples.




\bibliographystyle{mnras}
\bibliography{references} 

\begin{thebibliography}{}
\makeatletter
\relax
\def\mn@urlcharsother{\let\do\@makeother \do\$\do\&\do\#\do\^\do\_\do\%\do\~}
\def\mn@doi{\begingroup\mn@urlcharsother \@ifnextchar [ {\mn@doi@}
  {\mn@doi@[]}}
\def\mn@doi@[#1]#2{\def\@tempa{#1}\ifx\@tempa\@empty \href
  {http://dx.doi.org/#2} {doi:#2}\else \href {http://dx.doi.org/#2} {#1}\fi
  \endgroup}
\def\mn@eprint#1#2{\mn@eprint@#1:#2::\@nil}
\def\mn@eprint@arXiv#1{\href {http://arxiv.org/abs/#1} {{\tt arXiv:#1}}}
\def\mn@eprint@dblp#1{\href {http://dblp.uni-trier.de/rec/bibtex/#1.xml}
  {dblp:#1}}
\def\mn@eprint@#1:#2:#3:#4\@nil{\def\@tempa {#1}\def\@tempb {#2}\def\@tempc
  {#3}\ifx \@tempc \@empty \let \@tempc \@tempb \let \@tempb \@tempa \fi \ifx
  \@tempb \@empty \def\@tempb {arXiv}\fi \@ifundefined
  {mn@eprint@\@tempb}{\@tempb:\@tempc}{\expandafter \expandafter \csname
  mn@eprint@\@tempb\endcsname \expandafter{\@tempc}}}

\bibitem[\protect\citeauthoryear{{Ali} et~al.,}{{Ali} et~al.}{2015}]{Ali2015}
{Ali} Z.~S.,  et~al., 2015, \mn@doi [\apj] {10.1088/0004-637X/809/1/61}, \href
  {http://adsabs.harvard.edu/abs/2015ApJ...809...61A} {809, 61}

\bibitem[\protect\citeauthoryear{{Alonso}, {Bull}, {Ferreira}  \&
  {Santos}}{{Alonso} et~al.}{2015}]{Alonso2015}
{Alonso} D.,  {Bull} P.,  {Ferreira} P.~G.,   {Santos} M.~G.,  2015, \mn@doi
  [\mnras] {10.1093/mnras/stu2474}, \href
  {http://adsabs.harvard.edu/abs/2015MNRAS.447..400A} {447, 400}

\bibitem[\protect\citeauthoryear{{Anderson} et~al.,}{{Anderson}
  et~al.}{2018}]{Anderson2017}
{Anderson} C.~J.,  et~al., 2018, \mn@doi [\mnras] {10.1093/mnras/sty346}, \href
  {http://adsabs.harvard.edu/abs/2018MNRAS.tmp..340A} {}

\bibitem[\protect\citeauthoryear{{Battye}, {Browne}, {Dickinson}, {Heron},
  {Maffei}  \& {Pourtsidou}}{{Battye} et~al.}{2013}]{Battye2013}
{Battye} R.~A.,  {Browne} I.~W.~A.,  {Dickinson} C.,  {Heron} G.,  {Maffei} B.,
    {Pourtsidou} A.,  2013, \mn@doi [\mnras] {10.1093/mnras/stt1082}, \href
  {http://adsabs.harvard.edu/abs/2013MNRAS.434.1239B} {434, 1239}

\bibitem[\protect\citeauthoryear{{Bigot-Sazy} et~al.,}{{Bigot-Sazy}
  et~al.}{2015}]{BigotSazy2015}
{Bigot-Sazy} M.-A.,  et~al., 2015, \mn@doi [\mnras] {10.1093/mnras/stv2153},
  \href {http://adsabs.harvard.edu/abs/2015MNRAS.454.3240B} {454, 3240}

\bibitem[\protect\citeauthoryear{{Bigot-Sazy} et~al.,}{{Bigot-Sazy}
  et~al.}{2016}]{BigotSazy2016}
{Bigot-Sazy} M.-A.,  et~al., 2016, in {Qain} L.,  {Li} D.,  eds,  Astronomical
  Society of the Pacific Conference Series Vol. 502, Frontiers in Radio
  Astronomy and FAST Early Sciences Symposium 2015. p.~41 (\mn@eprint {arXiv}
  {1511.03006})

\bibitem[\protect\citeauthoryear{{Bull}, {Ferreira}, {Patel}  \&
  {Santos}}{{Bull} et~al.}{2015}]{Bull2015}
{Bull} P.,  {Ferreira} P.~G.,  {Patel} P.,   {Santos} M.~G.,  2015, \mn@doi
  [\apj] {10.1088/0004-637X/803/1/21}, \href
  {http://adsabs.harvard.edu/abs/2015ApJ...803...21B} {803, 21}

\bibitem[\protect\citeauthoryear{{Chang}, {Pen}, {Bandura}  \&
  {Peterson}}{{Chang} et~al.}{2010}]{Chang2010}
{Chang} T.-C.,  {Pen} U.-L.,  {Bandura} K.,   {Peterson} J.~B.,  2010, \mn@doi
  [\nat] {10.1038/nature09187}, \href
  {http://adsabs.harvard.edu/abs/2010Natur.466..463C} {466, 463}

\bibitem[\protect\citeauthoryear{{Chon}, {Challinor}, {Prunet}, {Hivon}  \&
  {Szapudi}}{{Chon} et~al.}{2004}]{Chon2004}
{Chon} G.,  {Challinor} A.,  {Prunet} S.,  {Hivon} E.,   {Szapudi} I.,  2004,
  \mn@doi [\mnras] {10.1111/j.1365-2966.2004.07737.x}, \href
  {http://adsabs.harvard.edu/abs/2004MNRAS.350..914C} {350, 914}

\bibitem[\protect\citeauthoryear{{Datta}, {Choudhury}  \& {Bharadwaj}}{{Datta}
  et~al.}{2007}]{Datta2007}
{Datta} K.~K.,  {Choudhury} T.~R.,   {Bharadwaj} S.,  2007, \mn@doi [\mnras]
  {10.1111/j.1365-2966.2007.11747.x}, \href
  {http://adsabs.harvard.edu/abs/2007MNRAS.378..119D} {378, 119}

\bibitem[\protect\citeauthoryear{Davidson, Lehmensiek, Young  et~al.}{Davidson
  et~al.}{2013}]{davidson2013}
Davidson D.,  Lehmensiek R.,  Young A.,   et~al., 2013, in Electromagnetics in
  Advanced Applications (ICEAA), 2013 International Conference on. pp
  1368--1371

\bibitem[\protect\citeauthoryear{Ellingson, Bunton  \& Bell}{Ellingson
  et~al.}{2001}]{ellingson2001}
Ellingson S.~W.,  Bunton J.~D.,   Bell J.~F.,  2001, The Astrophysical Journal
  Supplement Series, 135, 87

\bibitem[\protect\citeauthoryear{Ewall-Wice et~al.,}{Ewall-Wice
  et~al.}{2016}]{Ewall2016}
Ewall-Wice A.,  et~al., 2016, Monthly Notices of the Royal Astronomical
  Society, 460, 4320

\bibitem[\protect\citeauthoryear{Galt}{Galt}{1991}]{Galt1991}
Galt J.,  1991, in International Astronomical Union Colloquium. pp 213--221

\bibitem[\protect\citeauthoryear{Gao \& Enge}{Gao \& Enge}{2012}]{gao2012}
Gao G.~X.,  Enge P.,  2012, IEEE Transactions on aerospace and electronic
  Systems, 48, 2865

\bibitem[\protect\citeauthoryear{{Ghosh}, {Bharadwaj}, {Ali}  \&
  {Chengalur}}{{Ghosh} et~al.}{2011}]{Ghosh2011}
{Ghosh} A.,  {Bharadwaj} S.,  {Ali} S.~S.,   {Chengalur} J.~N.,  2011, \mn@doi
  [\mnras] {10.1111/j.1365-2966.2011.19649.x}, \href
  {http://adsabs.harvard.edu/abs/2011MNRAS.418.2584G} {418, 2584}

\bibitem[\protect\citeauthoryear{Green}{Green}{1985}]{green1985}
Green R.~M.,  1985, Spherical astronomy.
Cambridge University Press

\bibitem[\protect\citeauthoryear{{Hafez}, {Trojan}, {Albaqami}, {Almutairi},
  {Davies}, {Dickinson}  \& {Piccirillo}}{{Hafez} et~al.}{2014}]{Hafez2014}
{Hafez} Y.~A.,  {Trojan} L.,  {Albaqami} F.~H.,  {Almutairi} A.~Z.,  {Davies}
  R.~D.,  {Dickinson} C.,   {Piccirillo} L.,  2014, \mn@doi [\mnras]
  {10.1093/mnras/stt2476}, \href
  {http://adsabs.harvard.edu/abs/2014MNRAS.439.2271H} {439, 2271}

\bibitem[\protect\citeauthoryear{{Harper}, {Dickinson}, {Battye},
  {Roychowdhury}, {Browne}, {Ma}, {Olivari}  \& {Chen}}{{Harper}
  et~al.}{2018}]{Harper2017}
{Harper} S.,  {Dickinson} C.,  {Battye} R.~A.,  {Roychowdhury} S.,  {Browne}
  I.~W.~A.,  {Ma} Y.-Z.,  {Olivari} L.~C.,   {Chen} T.,  2018, \mn@doi [\mnras]
  {10.1093/mnras/sty1238}, \href
  {http://adsabs.harvard.edu/abs/2018MNRAS.tmp.1178H} {}

\bibitem[\protect\citeauthoryear{Hellbourg, Trainini, Weber, Moreau, Capdessus
  \& Boonstrd}{Hellbourg et~al.}{2012}]{hellbourg2012}
Hellbourg G.,  Trainini T.,  Weber R.,  Moreau E.,  Capdessus C.,   Boonstrd
  A.,  2012, in Signal Processing Conference (EUSIPCO), 2012 Proceedings of the
  20th European. pp 200--204

\bibitem[\protect\citeauthoryear{Hellbourg, Chippendale, Kesteven  \&
  Jeffs}{Hellbourg et~al.}{2014}]{hellbourg2014}
Hellbourg G.,  Chippendale A.,  Kesteven M.~J.,   Jeffs B.~D.,  2014, in Signal
  and Information Processing (GlobalSIP), 2014 IEEE Global Conference on. pp
  1286--1290

\bibitem[\protect\citeauthoryear{Hofmann-Wellenhof, Lichtenegger  \&
  Wasle}{Hofmann-Wellenhof et~al.}{2007}]{hofmann2007}
Hofmann-Wellenhof B.,  Lichtenegger H.,   Wasle E.,  2007, GNSS--global
  navigation satellite systems: GPS, GLONASS, Galileo, and more.
Springer Science \& Business Media

\bibitem[\protect\citeauthoryear{Lewis \& Bridle}{Lewis \&
  Bridle}{2002}]{Lewis2002}
Lewis A.,  Bridle S.,  2002, \mn@doi [Phys. Rev.] {10.1103/PhysRevD.66.103511},
  D66, 103511

\bibitem[\protect\citeauthoryear{Loeb \& Wyithe}{Loeb \&
  Wyithe}{2008}]{Loeb2008}
Loeb A.,  Wyithe J. S.~B.,  2008, Physical Review Letters, 100, 161301

\bibitem[\protect\citeauthoryear{Marquis \& Reigh}{Marquis \&
  Reigh}{2015}]{Marquis2015}
Marquis W.~A.,  Reigh D.~L.,  2015, Navigation, 62, 329

\bibitem[\protect\citeauthoryear{{Masui}, {McDonald}  \& {Pen}}{{Masui}
  et~al.}{2010}]{Masui2010}
{Masui} K.~W.,  {McDonald} P.,   {Pen} U.-L.,  2010, \mn@doi [\prd]
  {10.1103/PhysRevD.81.103527}, \href
  {http://adsabs.harvard.edu/abs/2010PhRvD..81j3527M} {81, 103527}

\bibitem[\protect\citeauthoryear{Masui et~al.,}{Masui et~al.}{2013}]{Masui2013}
Masui K.,  et~al., 2013, The Astrophysical Journal Letters, 763, L20

\bibitem[\protect\citeauthoryear{Montesano, Montesano, Caballero, Naranjo,
  Monjas, Cuesta, Zorrilla  \& Martinez}{Montesano
  et~al.}{2007}]{Montesano2007}
Montesano A.,  Montesano C.,  Caballero R.,  Naranjo M.,  Monjas F.,  Cuesta
  L.~E.,  Zorrilla P.,   Martinez L.,  2007, in The Second European Conference
  on Antennas and Propagation, EuCAP 2007. pp~1--6,
  \mn@doi{10.1049/ic.2007.1441}

\bibitem[\protect\citeauthoryear{{Olivari}, {Remazeilles}  \&
  {Dickinson}}{{Olivari} et~al.}{2016}]{Olivari2016}
{Olivari} L.~C.,  {Remazeilles} M.,   {Dickinson} C.,  2016, \mn@doi [\mnras]
  {10.1093/mnras/stv2884}, \href
  {http://adsabs.harvard.edu/abs/2016MNRAS.456.2749O} {456, 2749}

\bibitem[\protect\citeauthoryear{Patil et~al.,}{Patil et~al.}{2017}]{patil2017}
Patil A.,  et~al., 2017, The Astrophysical Journal, 838, 65

\bibitem[\protect\citeauthoryear{{Peebles}}{{Peebles}}{1973}]{Peebles1973}
{Peebles} P.~J.~E.,  1973, \mn@doi [\apj] {10.1086/152431}, \href
  {http://adsabs.harvard.edu/abs/1973ApJ...185..413P} {185, 413}

\bibitem[\protect\citeauthoryear{{Pen}, {Staveley-Smith}, {Peterson}  \&
  {Chang}}{{Pen} et~al.}{2009}]{Pen2009}
{Pen} U.-L.,  {Staveley-Smith} L.,  {Peterson} J.~B.,   {Chang} T.-C.,  2009,
  \mn@doi [\mnras] {10.1111/j.1745-3933.2008.00581.x}, \href
  {http://adsabs.harvard.edu/abs/2009MNRAS.394L...6P} {394, L6}

\bibitem[\protect\citeauthoryear{Pritchard \& Loeb}{Pritchard \&
  Loeb}{2008}]{Pritchard2008}
Pritchard J.~R.,  Loeb A.,  2008, Physical Review D, 78, 103511

\bibitem[\protect\citeauthoryear{{Santos} et~al.,}{{Santos}
  et~al.}{2015}]{Santos2015}
{Santos} M.,  et~al., 2015, Advancing Astrophysics with the Square Kilometre
  Array (AASKA14), \href {http://adsabs.harvard.edu/abs/2015aska.confE..19S}
  {p.~19}

\bibitem[\protect\citeauthoryear{{Santos} et~al.,}{{Santos}
  et~al.}{2017}]{Santos2017}
{Santos} M.~G.,  et~al., 2017, preprint, \href
  {http://adsabs.harvard.edu/abs/2017arXiv170906099S} {} (\mn@eprint {arXiv}
  {1709.06099})

\bibitem[\protect\citeauthoryear{{Shaw}, {Sigurdson}, {Pen}, {Stebbins}  \&
  {Sitwell}}{{Shaw} et~al.}{2014}]{Shaw2014}
{Shaw} J.~R.,  {Sigurdson} K.,  {Pen} U.-L.,  {Stebbins} A.,   {Sitwell} M.,
  2014, \mn@doi [\apj] {10.1088/0004-637X/781/2/57}, \href
  {http://adsabs.harvard.edu/abs/2014ApJ...781...57S} {781, 57}

\bibitem[\protect\citeauthoryear{Steigenberger, Thoelert  \&
  Montenbruck}{Steigenberger et~al.}{2017}]{Steigenberger2017}
Steigenberger P.,  Thoelert S.,   Montenbruck O.,  2017, Journal of Geodesy, pp
  1--16

\bibitem[\protect\citeauthoryear{{Switzer} et~al.,}{{Switzer}
  et~al.}{2013}]{Switzer2013}
{Switzer} E.~R.,  et~al., 2013, \mn@doi [\mnras] {10.1093/mnrasl/slt074}, \href
  {http://adsabs.harvard.edu/abs/2013MNRAS.434L..46S} {434, L46}

\bibitem[\protect\citeauthoryear{Union}{Union}{2004}]{ITU2004}
Union I.~T.,  2004, International Telecommunication Union radio regulations.,
  \url {www.itu.int}

\bibitem[\protect\citeauthoryear{Union}{Union}{2015}]{ITU2015}
Union I.~T.,  2015, Unwanted emissions in the out-of-band domain., \url
  {www.itu.int}

\bibitem[\protect\citeauthoryear{{Wilson}, {Rohlfs}  \&
  {H{\"u}ttemeister}}{{Wilson} et~al.}{2009}]{Wilson2009}
{Wilson} T.~L.,  {Rohlfs} K.,   {H{\"u}ttemeister} S.,  2009, {Tools of Radio
  Astronomy}.
Springer-Verlag, \mn@doi{10.1007/978-3-540-85122-6}

\bibitem[\protect\citeauthoryear{Wolz et~al.,}{Wolz et~al.}{2016}]{Wolz2016}
Wolz L.,  et~al., 2016, Monthly Notices of the Royal Astronomical Society, 464,
  4938

\makeatother
\end{thebibliography}




\bsp	
\label{lastpage}
\end{document}